\documentclass{appolb}
\usepackage{epsf,rotate}
\usepackage{amsmath,amstext,amsbsy,amsopn,amsfonts,amssymb}
\usepackage[dvips]{graphicx}
\usepackage[latin1]{inputenc}
\usepackage[english]{babel}
\usepackage[T1]{fontenc}
\usepackage{psfrag}

\newcommand{\eval}[2][\right]{\relax
  \ifx#1\right\relax \left.\fi#2#1\rvert}

\newcommand{\as}{\ifmmode\alpha_{\rm s}\else{$\alpha_{\rm s}$}\fi}
\newcommand{\asbar}{\ifmmode\wbar {\alpha}_{\rm s}\else{$\wbar{\alpha}_{\rm s}$}\fi}

 \catcode`\@=11

\def \lab #1 {\label{#1}}
\newcommand \ci [1] {\cite{#1}}

\def\r{{r}}
\def\c{{\hat{c}}}
\newcommand{\hhalf}{{\frac1{2}}}
\newcommand{\thalf}{{\frac3{2}}}
\newcommand{\fhalf}{{\frac5{2}}}
\newcommand{\shalf}{{\frac7{2}}}

\def\hph0{{\hphantom{0}}}
\newcommand{\cY}{{\cal Y}}
\newcommand{\cV}{{\cal V}}
\newcommand{\cS}{{\cal S}}

\begin{document}

\eqsec
\bibliographystyle{unsrt}

\title{Energy spectrum and wave-functions of 
four-dimensional 
Supersymmetric Yang-Mills Quantum Mechanics
for very high cut-offs}

\author{{\bf Jan~Kota{\'n}ski}
\address{M. Smoluchowski Institute of Physics, Jagellonian University\\
Reymonta 4, 30-059 Krak{\'o}w, Poland}}

%\date{April 15, 2004 }

\maketitle

\begin{abstract}
{\normalsize 
The spectrum of 
Supersymmetric Yang-Mills Quantum Mechanics
(SYMQM) in D=4 dimensions for $SU(2)$ gauge group
is computed for a maximal number of bosonic quanta $B\le60$ in 
the two-fermion
sector with the angular momentum $j=0$. We analyse the eigenfunctions
of discrete and continuous spectra, test the scaling relation
for the continuous spectrum
and confirm the dispersion relation
to high accuracy.}
\end{abstract}

\newline   

\vspace*{1cm}
\noindent TPJU-08/2006 \newline

\newpage 

\section{Introduction and summary of known results}

In this work we consider Supersymmetric Yang-Mills Quantum Mechanics
(SYMQM) \ci{Witten:1981nf,Claudson:1984th}, 
which in $D=10$ dimensions and for $SU(N \to \infty)$ gauge group
according to the remarkable hypothesis 
\ci{Banks:1996vh,Bigatti:1997jy,Taylor:2001vb}
is equivalent to $M-$theory 
of $D0$ branes. 
Despite the fact that the three loop calculations 
\ci{Dine:1998fu,Taylor:2001vb} question the exact 
equivalence, the SYMQM models possess
a lot of fascinating properties required in $M-$theory, \ie 
continuous  spectrum of scattering states with the threshold states
describing 
a supergravition \ci{Polchinski:1998rq}. 

The SYMQM emerges from the dimensional reduction of supersymmetric Yang-Mills 
field theory in the $D-1$ dimensional space to
the effective Quantum Mechanics of zero momentum modes in a single point 
\ci{Claudson:1984th,Halpern:1997fv}.
In a gluinoless sector
the energy spectrum of these system is identical to
the spectrum of $0-$volume glueballs 
\ci{Luscher:1982ma,Luscher:1983gm,vanBaal:1988qm,Wosiek:2002nm}.
Although the resulting models are much simpler than the 
original field theories,
they are rather complex with non-trivial solutions.
Researches on the systems with $D< 10$ for various gauge groups 
provide the global understanding of whole family. Thus, the SYMQM 
provides a simple laboratory to study many properties of 
supersymmetric systems 
\ci{Claudson:1984th,Halpern:1997fv}.
For $D=2$ and $SU(2)$ gauge group the system is exactly solvable. 
Going further to  $D=4$ dimensions
the model becomes non-trivial. It possesses both localized 
and non-localized states.

The simplest method of solving the energy eigenproblem for the 
$D=4$ model with $SU(2)$ group was introduced in 
Refs.~\ci{Wosiek:2002nm,Kotanski:2002fz}.
According to this method 
acting with bosonic and fermionic creation operators on 
the empty state
one can construct a basis of Fock space. Next, in order to find
the spectrum and the eigenfunctions of the Hamiltonian one can solve
the eigenproblem of the Hamiltonian matrix calculated in the constructed 
basis.
Since we have an infinite number of the states in the Fock space in order
to perform numerical calculation we have to cut some  basis vectors off.
Therefore, in the above method one considers only Fock states with a number of 
bosonic quanta $n_B$ smaller or equal to cut-off $B$.
At the end making analysis of the spectrum as a function of cut-off $B$
we are able to find the energy
for the model without a cut-off.
It turns out that   due to commutativity properties between the Hamiltonian and
the operator of number of fermionic quanta, $n_F$, the  system separates into
sectors with determined $n_F=0,1,\ldots,6$. Thus,  we can solve 
the eigenproblem in smaller Fock basis with fixed $n_F$.
The above method allowed to calculate
the spectrum in all fermionic sectors with an arbitrary 
angular momentum $j$ for 
the cut-off $B=8$ and the Witten index for the model 
\ci{Wosiek:2002nm,Kotanski:2002fz}.

It turns out that the Hamiltonian also commutes with an operator of the 
angular momentum $j$. This makes possible to construct the basis 
with fixed both $n_F$ and the angular momentum, $j$.
It was done by van Baal in Refs.~\ci{Koller:1987fq,vanBaal:2001ac} for
$n_F=0,2,4,6$ and $j=0$ but with the cut-off $B=39$.
Calculations for other sectors are more complicated
within this approach.

The aim of this work is to analyse the spectrum
of the $D=4$ model with $SU(2)$ gauge group more precisely
by enlarging cut-off yet further.
Using the van Baal's method we concentrate on the sector with $F=2$
and $j=0$ where the supersymmetric vacuum state appears to be 
\ci{vanBaal:2001ac}.
In this sector the eigenvalues of the Hamiltonian form discrete as well
as continuous spectrum \ci{deWit:1988ct,Nicolai:1998ic}. 
For high cut-offs the wave-functions of discrete states
converge with $B$ 
and therefore we are able to
build these eigenstates 
and describe properties of their wave-functions.
Moreover,
we can test the scaling properties \ci{Trzetrzelewski:2003sz}
appearing for the continuous spectrum.

In the beginning of 
the next section
%this paper 
we introduce a notation following
Ref.~\ci{vanBaal:2001ac} and describe the SYMQM model and show
how one can construct Fock space in the $F=2$ an d $j=0$ sector
and calculate the Hamiltonian matrix.
% \ci{vanBaal:2001ac}. 
We also present the construction of the eigenstates of the Hamiltonian.
Next we show the results which include the energy spectrum of the model
for cut-off $B\le 60$. Moreover, we construct and describe properties of 
the eigenfunctions for the discrete and continuous spectrum. 
At the end we confirm the scaling relation  
for the continuous spectrum which was derived 
in Ref.~\ci{Trzetrzelewski:2003sz}
for the cut free-particle and only
conjectured for interaction systems.

%%%%%%%%%%%%%%%%%%%%%%%%%%%%%%%%%%%%%%%%%%%%%%%%%%%%%%%%%%%%%%%%%%%%%%%%
\section{Hamiltonian eigenproblem}
\subsection{Definitions}

Following Refs.~\ci{Luscher:1982ma,vanBaal:2001ac} the supersymmetric Hamiltonian
of Yang-Mills quantum mechanics with the $SU(2)$ gauge group
has a form
\begin{equation}
H \delta_{\alpha \beta}= 
\hhalf  \{Q_{\alpha},Q_{\beta}^{\dagger } \}\,,
\lab{eq:HQQ}
\end{equation}
where $Q$ and $Q^{\dagger}$ are the generators of
SUSY in the coordinate representation of the field theory defined as
\begin{equation}
Q_{\alpha }=\sigma ^{j}_{\alpha \dot{\beta }}
\bar{\lambda }_{a}^{\dot{\beta }}
\left( -i\frac{\partial }{\partial V_{a}^{j}}-iB_{j}^{a}\right)
\qquad
\bar{Q}_{\dot{\alpha }}=\lambda _{a}^{\beta }
\sigma ^{j}_{\beta \dot{\alpha }}
\left( -i\frac{\partial }{\partial V_{a}^{j}}+iB_{j}^{a}\right)\,,
\lab{eq:QQ}
\end{equation}
with Pauli matrices $ \sigma ^{j}=\tau^j$
and Weyl spinors $\lambda _{a}^{\beta }$.
Here, a {\it colour magnetic} field of $SU(2)$
is defined as
\begin{equation}
B_{i}^{a}=-\frac{1}{2}g\varepsilon _{ijk}
\varepsilon _{abc}V_{j}^{b}V_{k}^{c}\,,
\lab{eq:BVV}
\end{equation}
where $V_j^c$ are bosonic variables 
with colour indices $c=1,2,3$ and spatial indices $j=1,2,3$.
The Weyl spinors  $\lambda _{a}^{\beta }$
with $\beta=1,2$ spinor indices and $a=1,2,3$ colour indices 
satisfy the following 
anti-commutation relations
\begin{equation}
\{\lambda ^{a\alpha },
\bar{\lambda }^{b\dot{\beta }}\}
=\bar{\sigma }^{\dot{\beta }\alpha }_{0}\delta ^{ab}\,,
\quad
\{\lambda ^{a\alpha },\lambda ^{b\beta }\}=0\,,
\quad
\{\bar{\lambda }^{a\dot{\alpha }},
\bar{\lambda }^{b\dot{\beta }}\}=0\,,
\lab{eq:comLL}
\end{equation}
where $\sigma^0$ is the unit matrix.

The SUSY generators satisfy
\begin{equation}
\{Q_{\alpha },\bar{Q}_{\dot{\alpha }}\}=
2(\sigma _{0})_{\alpha \dot{\alpha }}{
\mathcal{H}}-2(\sigma ^{i})_{\alpha \dot{\alpha }}V_{i}^{a}{\mathcal{G}}_{a}\,,
\lab{eq:comQQ}
\end{equation}
where 
\begin{equation}
{\mathcal{G}}_{a}=ig\varepsilon _{abc}
\left( V_{j}^{c}\frac{\partial }{\partial V_{j}^{b}}-\bar{\lambda }^{b}
\bar{\sigma }_{0}\lambda ^{c}\right)\,,
\lab{eq:Ga}
\end{equation}
is the generator of the $SU(2)$ gauge transformation
while the Hamiltonian density \ci{Halpern:1997fv} is defined by
\begin{equation}
{\mathcal{H}}
=-\frac{1}{2}\frac{\partial ^{2}}{\partial V_{i}^{a}\partial V_{i}^{a}}
+\frac{1}{2}B_{i}^{a}B_{i}^{a}
-ig\varepsilon _{abc}\bar{\lambda }^{a}\bar{\sigma }^{j}
\lambda^{b}V_{j}^{c}\,.
\lab{eq:Hden}
\end{equation}
One may rescale bosonic variables $V_i^a$
introducing new variables $\hat{c}_i^{a}$:
\begin{equation}
V_i^a=\frac{1}{g^{1/3}(L)\,L}\hat{c}_{i}^{a}\,.
\lab{eq:scal}
\end{equation}
Thus, performing the approximation
of constant fields with $g(L\to0) \to 0$ 
\ci{Luscher:1982ma} 
%\ci{vanBaal:2001ac} 
with $g=g(L)\,L^3$
we
obtain a Hamiltonian $H$ independent of the coupling constant $g$:
\begin{equation}
\int d^{3} x {\mathcal{H}} \equiv g^{2/3}(L)H/L,
\lab{eq:HHden}
\end{equation}
where
\begin{equation}
H=H_B+H_F,
\lab{eq:HBF}
\end{equation}
with the bosonic part
\begin{equation}
H_{B}
=-\frac{1}{2}\left( \frac{\partial }{\partial \hat{c}_{i}^{a}}\right) ^{2}
+\frac{1}{2}\left( \hat{B}_{i}^{a}\right) ^{2},
\lab{eq:HB}
\end{equation}
and the fermionic one
\begin{equation}
H_{F}=-i\varepsilon _{abd}\bar{\lambda }^{a}
\bar{\sigma }^{i}\lambda ^{b}\hat{c}_{i}^{d}\,.
\lab{eq:HF}
\end{equation}
Here, the {\it colour magnetic} field is rescaled to
\begin{equation}
\hat{B}_{i}^{a}=-\frac{1}{2}\varepsilon _{ijk}
\varepsilon _{abd} \c_{j}^{b} \c_{k}^{d}\,.
\lab{eq:hB}
\end{equation}

\subsection{A straightforward approach}
In this model 
there are 6 fermionic degrees of freedom, $\lambda_{a}^{\beta}$.
An operator
which gives the number of fermions, $n_F$,
commutes with the Hamiltonian (\ref{eq:HBF}).
Thus, our system splits into $7$ sectors
enumerated by $n_F$ values. Since the sectors
are related by a particle-hole symmetry
\begin{equation}
n_F \leftrightarrow 6 - n_F\,,
\lab{eq:PHsym}
\end{equation}
we have four independent sectors described by $n_F=0,1,2,3$.

In order to solve the eigenequation of the Hamiltonian (\ref{eq:HBF})
we construct an infinite basis of the Fock space
by acting with 
$\c_k^d$ and ${\bar \lambda}_k^{\dot{\alpha}}$
on the empty state:
\begin{equation}
|n \rangle =
\sum_{\scriptsize
\begin{array}{c}
\rm{contractions}\\
\{a_1,\ldots,a_r\}
\end{array}}
\c_{k_1}^{a_1} \ldots \c_{k_m}^{a_m} 
{\bar \lambda}_{a_{m+1}}^{\dot{\alpha}} \ldots {\bar \lambda}_{a_r}^{\dot{\beta}} 
|0 \rangle\,,
\lab{eq:state}
\end{equation}
where $\c_k^d$ and ${\bar \lambda}_k^{\dot{\alpha}}$
are given by appropriate linear combinations of the creation and annihilation
bosonic and fermionic operators, respectively,
while
the sum goes over gauge invariant linear combinations
of $\c_k^d$ and ${\bar \lambda}_k^{\dot{\alpha}}$, 
\ci{Wosiek:2002nm}\footnote{Here, calculations are performed 
in Fock-space representation}.
Next, we act with (\ref{eq:HBF}) on the basis states (\ref{eq:state})
calculating matrix elements of the Hamiltonian. Finally, 
we diagonalize the Hamiltonian matrix finding eigenvalues and
eigenstates of (\ref{eq:HBF}).

Contrary to fermions we have an infinite number of bosons, 
$n_B=0,\ldots,\infty$.
Moreover, the operator which describes a number of bosons $n_B$
does not commute with the Hamiltonian (\ref{eq:HBF}).
Therefore, our Hamiltonian matrix is infinite.
To simplify the problem we cut the basis (\ref{eq:state})
by considering only the states with $n_B \le B$, where 
$B$ is defined as a cut-off. Next, analysing the limit 
$B \to \infty$ we recover the spectrum of the full, infinite
Hamiltonian matrix.

\begin{table}[ht!]
\begin{center}
\begin{tabular}{cccccc} 
\hline\hline
  $n_F$ &  $ 0 $ & $ 1 $ & $ 2 $ & $ 3 $  & \\  \hline
  $n_B$ &  $ N_{s}$\ \ \ $  \Sigma_0$ \ \  & $ N_{s} \ \ \  \Sigma_1$ \ \  &  $ N_{s} \ \ \  \Sigma_2$ \ \  & $ N_{s}\ \ \   \Sigma_3$  \ \ &  $\Sigma_B - \Sigma_F$ \\
   \hline
  0 &    1 \     1 \ \   &  -  \    -   \  &   1 \     1  \  &    4 \     4  \ & 0  \\
  1 &   -  \     1 \ \   &   6 \     6  \  &   9 \    10  \  &    6 \    10  \ & 0  \\
  2 &    6 \     7 \ \   &   6 \    12  \  &  21 \    31  \  &   42 \    52  \ & 0  \\
  3 &    1 \     8 \ \   &  36 \    48  \  &  63 \    94  \  &   56 \   108  \ & 0  \\
  4 &   21 \    29 \ \   &  36 \    84  \  & 111 \   205  \  &  192 \   300  \ & 0  \\
  5 &    6 \    35 \ \   & 126 \   210  \  & 240 \   445  \  &  240 \   540  \ & 0  \\
  6 &   56 \    91 \ \   & 126 \   336  \  & 370 \   815  \  &  600 \  1140  \ & 0  \\
  7 &   21 \   112 \ \   & 336 \   672  \  & 675 \  1490  \  &  720 \  1860  \ & 0  \\
  8 &  126 \   238 \ \   & 336 \  1008  \  & 960 \  2450  \  & 1500 \  3360  \ & 0  \\
\hline
 $j_{\max}$ &  8          &    17/2         &     9           &   19/2         &    \\
   \hline\hline
\end{tabular}
\end{center}
\caption{Size of the basis generated in
specified fermionic sectors $n_F$ from Ref.~\ci{Kotanski:2002fz}}
\lab{tab:Enf}
\end{table}

Eigenstates of the model are 
linear combinations of basis states (\ref{eq:state}).
It turns out that for given $n_B$ we have a large number of states.
Even if we divide the problem into the separate fermionic sectors,
the separate bases have large numbers of states.
One can see it in
Table \ref{tab:Enf}  from 
Ref.~\ci{Kotanski:2002fz}
where the authors show the size of the basis containing states
generated in specified fermionic sectors labelled  by $n_F$ 
and for given $n_B$.
In this Table $N_s$ describes a number of
basis vectors
with a given bosonic quantum numbers whereas
$\Sigma$ is a sum of vectors in the basis
with maximal quantum number $n_B$.
The last column shows that the number of bosonic states
$\Sigma_B=\Sigma_0+\Sigma_2+\Sigma_4+\Sigma_6$
is equal to the number of fermionic states 
$\Sigma_F=\Sigma_1+\Sigma_3+\Sigma_5$
as required by supersymmetry. 

\begin{figure}[ht!]
\centerline{
\psfrag{B}{$B$}
\psfrag{F}[rc][bc]{$n_F$}
\psfrag{E}[rc][bc]{$E$}
\epsfysize 9cm \epsfbox{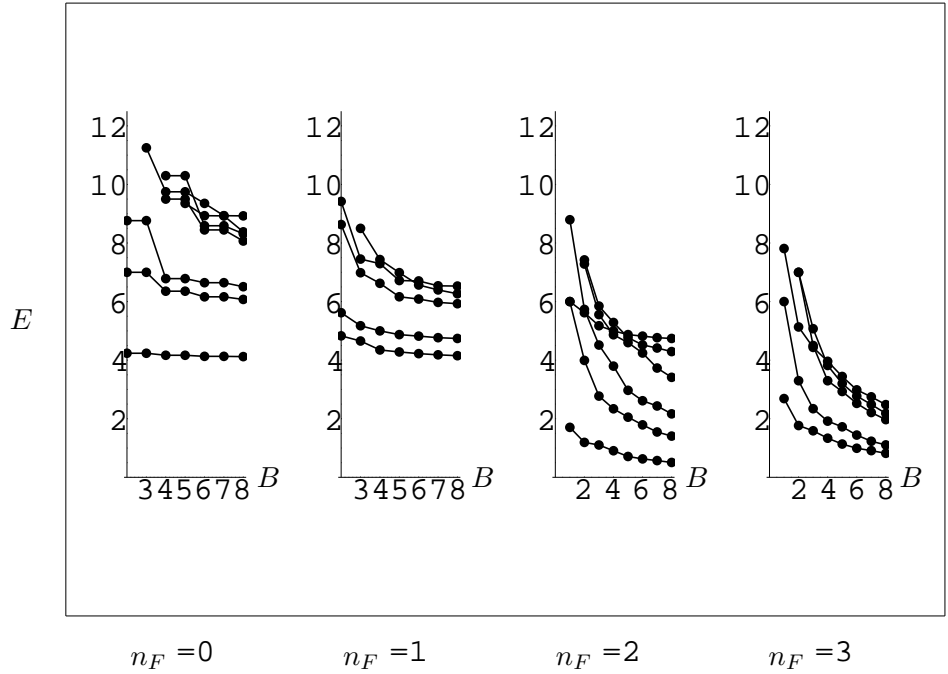}
}
\caption{Dependence of the energy $E$ on a 
a cut-off $B\ge n_B$ in sectors with various 
$n_F$ 
obtained in Ref.~\ci{Kotanski:2002fz}}
\lab{fig:Enbnf}
\end{figure}

In Fig.~\ref{fig:Enbnf} 
we present
a dependence of the energy on the cut-off.
One can see that for this large number of
bosonic quanta the energy converges
to the constant value rather slowly, especially for 
$n_F=2,3,4$. Thus, in order to
simplify this problem, we find another operator which commutes
with the Hamiltonian, \ie the total angular momentum $j$.
Using the operator describing a fermion number, 
$n_F$, and the total angular momentum, we can divide the spectrum
into smaller sectors labelled by $n_F$ and $j$.
This will be discussed below.

\subsection{Analytic change of variables}

In the rest of the paper we consider the case with 
$n_F=2$ and $j=0$. This case is most interesting
because it contains the vacuum state of the system.
Moreover, the energy has discrete and continuous spectrum.
We will use the notation following
the work of van Baal \ci{vanBaal:2001ac,Koller:1987fq}. 

The gauge invariance and the vanishing angular momentum $j$
allow us to reduce a number of bosonic variables \ci{vanBaal:2001ac}.
The reduction can be rewritten as a 
diagonalization of $\c_i^a$:
\begin{equation}
\c_{i}^{a}=\sum _{j=1}^3 R_{ij}x_{j}T^{ja}\,,
\lab{eq:cdiag}
\end{equation}
where matrices $R,T\in SO(3)$. Thus, from nine bosonic variables 
$\c_{i}^{a}$ one obtains three invariants 
of both gauge and rotation groups:
\begin{equation}
-\infty < x_i < \infty\,,
\quad
\mbox{where}
\quad
i = 1,2,3\,.
\lab{eq:xi}
\end{equation}
Additionally, the second set of invariant variables 
$(\r,u,v)$ is introduced where
\begin{equation}
\r^{2}=(\hat{c}_{j}^{a})^{2}
=\sum _{j=1}^3 x_{j}^{2}\,,
\lab{eq:rcrd}
\end{equation}
defines the
radius $\r$ 
in the $\{x_j\}-$space, so that  $0\le \r \le \infty$,
while
\begin{equation}
u=\r^{-4}(\hat{B}_{j}^{a})^{2}
=\r^{-4}\sum_{i>j=1}^3 x_{i}^{2}x_{j}^{2}\,,
\lab{eq:ucrd}
\end{equation}
takes values $0\le u \le \frac{1}{3}$
and corresponds to the rescaled bosonic potential 
\begin{equation}
V(\vec{x})=x_1^2 x_2^2+x_1^2 x_3^2+x_2^2 x_3^2=u\, \r^4\,.
\lab{eq:Vx}
\end{equation}
An example of equipotential surface is shown in Fig.~\ref{fig:epot}.
\begin{figure}[ht!]
\centerline{
\begin{picture}(90,90)
\put(0,0){\epsfysize9cm \epsfbox{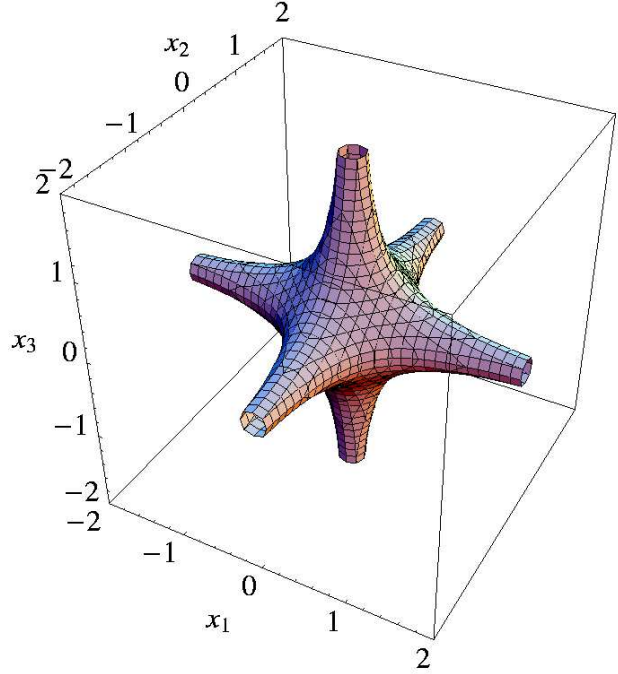}}
\end{picture}
}
\caption{An example of equipotential surface for 
$V(\vec{x})=u\, \r^4=x_1^2 x_2^2+x_1^2 x_3^2+x_2^2 x_3^2=0.1$}
\lab{fig:epot}
\end{figure}
The potential has minimum on $x_i-$axes.
Due to the rescaling property (\ref{eq:Vx}) of $V(\vec x)$
all equipotential surfaces with different values of energy
have the same shape.

It turns out that for $n_F=2$ the fermionic part 
of the Hamiltonian
breaks the symmetry 
\begin{equation}
x_1 x_2 x_3 \leftrightarrow - x_1 x_2 x_3\,.
\lab{eq:xtmx}
\end{equation}
Thus, in order to describe
the system completely one has to introduce the remaining variable
\begin{equation}
v=\r^{-3}\det \hat{c}
=\r^{-3}\prod _{j=1}^{3} x_{j}\,,
\lab{eq:vcrd}
\end{equation}
which may take positive as well as negative values 
$-1/\sqrt{27}\le v \le 1/\sqrt{27}$.

One can notice that the transition from $(\r,u,v)$ to $(x_1,x_2,x_3)$
is not unique. The transformation functions $x_j(\r,u,v)$ 
between coordinates are 24-valued functions and we can choose one branch,
\ie imposing $|x_1|\le x_2 \le x_3$.
It turns out that for 
$n_F=2$ and $j=0$ 
bosonic terms of our calculations 
can be rewritten in $(x_1,x_2,x_3)$ variables or 
equivalently in $(\r,u,v)$ coordinates.
After integrating out
six angular variables,
the integral measure changes into
\begin{equation}
d^{9}\hat{c}=\frac{2}{3}\pi ^{4}Jd^{3}x\,,
\lab{eq:d9c}
\end{equation}
with the Jacobian $J$ defined as
\begin{equation}
J\equiv \prod _{i>j}|x_{i}^{2}-x_{j}^{2}|
=\r^{6}\sqrt{u^{2}(1-4u)-v^{2}(4-18u+27v^{2})}\,.
\lab{eq:Jac}
\end{equation}
 
Let us consider the basis vectors (\ref{eq:state}).
For $n_F=2$ they contain two fermionic creation operators 
$\bar \lambda_{\dot{\alpha }}^{a}$.
We can form  either antisymmetric
\begin{equation}
|{\mathcal{V}}\rangle \equiv 
{{\mathcal{V}}_{j}}^{c}{{\mathcal{I}}^{j}}_{c}=-2i{{\mathcal{V}}_{j}}^{c}
\varepsilon _{abc}
\bar{\lambda }_{\dot{\alpha }}^{a}
{(\bar{\sigma }^{j0})^{\dot{\alpha }}}_{\dot{\beta }}
\bar{\lambda }^{b\dot{\beta }}|0\rangle\,,
\lab{eq:PsiV}
\end{equation}
or symmetric
\begin{equation}
|{\mathcal{S}}\rangle \equiv 
{\mathcal{S}}_{ab}{\mathcal{J}}^{ab}=-{\mathcal{S}}_{ab}
\bar{\lambda }_{\dot{\alpha }}^{a}\bar{\lambda }_{\dot{\beta}}^{b}
\epsilon ^{\dot{\alpha }\dot{\beta }}|0\rangle \,,
\lab{eq:PsiS}
\end{equation}
combinations,
where ${{\mathcal{V}}_{j}}^{c}$ and ${\mathcal{S}}_{ab}$
are antisymmetric and symmetric combinations of bosonic variables 
$\c_j^a$ respectively, while $\bar{\sigma }^{j0}$ is 
related to Pauli matrices by
$\bar{\sigma }^{j0}=\hhalf \tau_j$.
The covariance gives us their structure as 
\begin{equation}
{{\mathcal{V}}_{j}}^{a}=h_{1}(\r,u,v)\hat{c}_{j}^{a}/\r
-h_{2}(\r,u,v)\hat{B}_{j}^{a}/\r^{2}
+h_{3}(\r,u,v)\hat{c}_{j}^{b}\hat{c}_{k}^{b}\hat{c}_{k}^{a}/\r^{3}\,,
\lab{eq:Vja}
\end{equation}
and
\begin{equation}
{\mathcal{S}}^{ab}=h_{4}(\r,u,v)\delta ^{ab}
-h_{5}(\r,u,v)\hat{c}_{j}^{a}\hat{c}_{j}^{b}/\r^{2}
+h_{6}(\r,u,v)
\hat{c}_{j}^{b}\hat{c}_{j}^{d}\hat{c}_{k}^{d}\hat{c}_{k}^{b}/\r^{4}\,,
\lab{eq:Sab}
\end{equation}
where $h_i(\r,u,v)$ are functions of invariant variables.
Finally, our basis vector may have both invariant parts, so that
\begin{equation}
|\Psi \rangle= |{\mathcal{V}}\rangle+|{\mathcal{S}}\rangle\,.
\lab{eq:PsiVS}
\end{equation}

The Eqs.~(\ref{eq:Vja})-(\ref{eq:PsiVS}) can be rewritten as
\begin{equation}
|\Psi \rangle =\sum _{\mu=1}^{6}h_{\mu}(\r,u,v)|e_{\mu}(u,v)\rangle\,,
\lab{eq:Psih}
\end{equation}
where
\begin{equation}
\begin{array}{ll}
|e_{1}(u,v)\rangle=\hat{c}_{j}^{a}/\r {{\mathcal{I}}^{j}}_{a}\,, &
|e_{4}(u,v)\rangle=\delta ^{ab} {\mathcal{J}}_{ab}\,, 
\vspace{0.2cm}\\
|e_{2}(u,v)\rangle=\hat{B}_{j}^{a}/\r^{2} {{\mathcal{I}}^{j}}_{a}\,, &
|e_{5}(u,v)\rangle=\hat{c}_{j}^{a}\hat{c}_{j}^{b}/\r^{2} 
{\mathcal{J}}_{ab}\,, 
\vspace{0.2cm}\\
|e_{3}(u,v)\rangle=\hat{c}_{j}^{b}\hat{c}_{k}^{b}\hat{c}_{k}^{a}/\r^{3}
{{\mathcal{I}}^{j}}_{a}\,, \quad &
|e_{6}(u,v)\rangle=\hat{c}_{j}^{b}\hat{c}_{j}^{d}\hat{c}_{k}^{d}\hat{c}_{k}^{b}/\r^{4} {\mathcal{J}}_{ab}\,.
\end{array}
\end{equation}
Lowering of indices can be done with 
$\epsilon_{\alpha \beta}=\epsilon_{\dot{\alpha} \dot{\beta}}=
-i \tau_2$ for spinor indices,
$\delta_{ab}$ for colour group indices and
$\delta_{ij}$
for 
space
indices.
Now, we are ready to reformulate our Hamiltonian (\ref{eq:HBF}) 
in terms of $(\r,u,v)$.

The fermionic part of $H_F$
can be expanded in the space spanned by 
$|e_{\mu} \rangle \equiv |e_{\mu}(u,v)\rangle$, \ie
\begin{equation}
H_{F}|e_{\mu}\rangle 
=\sum _{\nu=1}^{6}|e_{\nu}\rangle H_{F}^{\nu \mu}\,,
\lab{eq:HFe}
\end{equation}
which gives
a $6 \times 6$ matrix in a form
\begin{equation}
H_{F}^{\mu \nu}/\r= \left(
\begin{array}{cccccc}
0 & 1 & -v & 2 & 1 & 1-u\\
2 & 0 & 1 & 0 & 0 & -v\\
0 & -1 & 0 & 0 & -1 & -1\\
2 & 4v & 2-4u & 0 & 0 & 0\\
-2 & 0 & 0 & 0 & 0 & 0\\
0 & 0 & -2 & 0 & 0 & 0
\end{array}
\right)\,.
\lab{eq:HFmat}
\end{equation}
The vectors $|e_{\mu}(u,v)\rangle$ are not orthogonal
and their scalar products\footnote{These scalar products, 
\ie $\langle \cdot | \cdot \rangle_F$, are performed with integration
only over fermionic degrees of freedom}
can be written in a matrix form as
\begin{equation}
N^{\mu \nu}=\langle e_{\mu}|e_{\nu}\rangle_F=
\left( 
\begin{array}{cccccc}
8 & 24v & 8X & 0 & 0 & 0\\
24v & 8u & 8v & 0 & 0 & 0\\
8X & 8v & 8Y & 0 & 0 & 0\\
0 & 0 & 0 & 12 & 4 & 4X\\
0 & 0 & 0 & 4 & 4X & 4Y\\
0 & 0 & 0 & 4X & 4Y & 4Z
\end{array}
\right)\,,
\lab{eq:Nnm}
\end{equation}
where 
\begin{equation}
X=1-2u, 
\quad
Y=1-3u+3v^{2}
\quad
\mbox{and}  
\quad
Z=1+2u^{2}-4u+4v^{2}.
\lab{eq:XYZ}
\end{equation}
Thus, calculating scalar products of the states (\ref{eq:Psih})
we have to use the $N^{\nu \mu}$ matrix, \ie 
\begin{equation}
\langle \Psi |\Psi '\rangle 
=\int d^{9}\hat{c}\sum _{\mu,\nu=1}^{6}h^{\ast }_{\mu}N^{\mu \nu}h'_{\nu}\,,
\lab{eq:PsiPsi}
\end{equation}
with $h_{\nu}$ defined in (\ref{eq:Vja})-(\ref{eq:Sab}).
Similarly, the matrix elements have a form
\begin{equation}
\langle \Psi |H|\Psi '\rangle 
=\int d^{9}\hat{c}\sum _{\mu,\nu,p=1}^{6}h^{\ast }_{\mu}N^{\mu \nu}
H^{\nu \rho}h'_{\rho}\,.
\lab{eq:PsiHPsi}
\end{equation}

The kinetic part of the Hamiltonian (\ref{eq:HBF}) is given by
\begin{equation}
-\hhalf\frac{\partial^2}{(\partial\c_i^a)^2}=-\hhalf J^{-1}(\vec x)
\frac{\partial}{\partial x_j}J(\vec x)\frac{\partial}{\partial x_j}=
-\hhalf\left(\r^{-8}\frac{\partial}{\partial\r}\r^8\frac{\partial}{\partial\r}
+\frac{\Delta(u,v)}{\r^2}\right)\,,
\lab{eq:Hkin}
\end{equation}
where the Laplacian on a $9-$dimensional sphere reads
\begin{eqnarray}
\Delta(u,v)&=&4(3v^2+u-4u^2)\frac{\partial^2}{(\partial u)^2}
              +8(1-3u)v\frac{\partial^2}{\partial u\partial v}
              +(u-9v^2)\frac{\partial^2}{(\partial v)^2}\nonumber\\
            &&+4(2-11u)\frac{\partial}{\partial u}
              -30v\frac{\partial}{\partial v}.
\lab{eq:del}
\end{eqnarray}

Similarly to (\ref{eq:HFe}), the matrix elements of $\Delta(u,v)$
can be calculated as
\begin{equation}
-\hhalf\Delta(u,v)\sum_{\mu=1}^6h_{\mu}(\r,u,v)|e_{\mu}(u,v)\rangle=
\sum_{\mu,\nu=1}^6|e_{\nu}(u,v)\rangle \hat H_\Delta^{\nu \mu}
h_{\mu}(\r,u,v)\,,
\lab{eq:HD}
\end{equation}
where $\hat H_\Delta^{\nu \mu}$ is found in Ref.~\ci{vanBaal:2001ac} as
\begin{equation}
\hat H^{\mu \nu}_\Delta=-\hhalf\delta^{\mu \nu}\Delta(u,v)-\hhalf
\left( 
\begin{array}{cc}
\Delta^1_\cV&\oslash\\
\oslash&\Delta^1_\cS\\
\end{array}
\right)
-\hhalf
\left( 
\begin{array}{cc}
\Delta^0_\cV&\oslash\\
\oslash&\Delta^0_\cS\\
\end{array}
\right)\,,
\lab{eq:Hdel}
\end{equation}
with
\begin{equation}
\Delta^1_\cV\equiv2
\left( 
\begin{array}{ccc}
(2-4u)\partial_u-3v\partial_v&\partial_v+
                       2v\partial_u&3v\partial_v+6u\partial_u\cr
       \partial_v&(2-8u)\partial_u-6v\partial_v&-6v\partial_u\cr
         -2\partial_u&-\partial_v&-12u\partial_u-9v\partial_v\cr
\end{array}
\right),
\end{equation}
\begin{equation}
\Delta^1_\cS\equiv2
\left( 
\begin{array}{ccc}
0&2v\partial_v&-8v^2\partial_u\cr
    0&(4-8u)\partial_u-6v\partial_v&4v\partial_v+8u\partial_u\cr
                   0&-4\partial_u&-16u\partial_u-12v\partial_v\cr
\end{array}
\right),
\end{equation}
and
\begin{equation}
\Delta^0_\cV\equiv2
\left( 
\begin{array}{ccc}
\!-4&~0&~7\cr~0&\!-9&~0\cr~0&~0&\!-15\cr
\end{array}
\right),
\quad
\Delta^0_\cS\equiv
\left( 
\begin{array}{ccc}
~0&~3&~1\cr~0&\!-9&~11\cr~0&~0&\!-22\cr
\end{array}
\right).
\end{equation}

To sum up, we obtain the $6 \times 6$ Hamiltonian matrix in a form
\begin{equation}
\hat{H}^{\mu \nu}=
-\frac{1}{2}\delta ^{\mu \nu}\r^{-8}
\frac{\partial }{\partial \r}\r^{8}\frac{\partial }{\partial \r}
+\r^{-2}\hat{H}_{\Delta }^{\mu \nu}
+\frac{1}{2}\delta ^{\mu \nu}\r^{4}u
+\r\hat{H}_{F}^{\mu \nu}\,,
\lab{eq:Hmn}
\end{equation}
where
$\hat{H}_{F}^{\mu \nu}=H_{F}^{\mu \nu}/\r$.

\subsection{The Fock space}

Following Ref. \ci{vanBaal:2001ac} 
we choose 
the eigenvectors of the harmonic oscillator 
as the basis vectors (\ref{eq:Psih}).
Substituting 
$(\r,u,v)$ variables these eigenfunctions
separate into a spherical
and radial part as follows\footnote{In the following sections 
all sums will be written directly. There is no summation over repeating 
indices assumed a priori}
\begin{equation}
|\hat{\Psi}^{(n,\ell,m)}(\r,u,v)\rangle=
\sum_{\mu=1}^6 h_{\mu}^{(n,\ell,m)}(\r,u,v)|e_{\mu} \rangle \,,
\lab{eq:hPsih2}
\end{equation}
with
\begin{equation}
h_{\mu}^{(n,\ell,m)}(\r,u,v)=
\cY_{\mu}^{(\ell,m)}(u,v)\phi_n^{\ell}(\r)\,.
\lab{eq:hYphi}
\end{equation}
Next, to ortonormalize the basis (\ref{eq:hPsih2})
the Gram-Schmidt process 
\begin{equation}
|\hat{\Psi}^{(n,\ell,m)}(\r,u,v)\rangle 
\stackrel{\rm ortonorm.}{\longrightarrow}
|{\Psi}^{(n,\ell,m)}(\r,u,v)\rangle\,,
\lab{eq:Psih2}
\end{equation}
is performed.

\begin{table}[ht!]
\begin{center}
\begin{tabular}{|c|c|l|}
\hline 
$\rule[-2mm]{0cm}{7mm}$
$
\hhalf(\ell-3)$ & $m$ &$\cY^{(\ell,m)}=

(\cY^{(\ell,m)}_1,\cY^{(\ell,m)}_2,\ldots,\cY^{(\ell,m)}_6)$\\
\hline
\hline
$\rule[-2mm]{0cm}{7mm}$
$0$&$0$&$
(0,0,0,1,0,0)\,
            \sqrt{35/2}/8\pi^2$\\ 
\hline
$\rule[-2mm]{0cm}{7mm}$
$\hhalf$&$\hhalf$&$
(1,0,0,0,0,0)\,
            \sqrt{105}/16\pi^2$\\ 
\hline
$\rule[-0mm]{0cm}{5mm}$
$1$&$0$&$
(0,1,0,0,0,0)\,
            \sqrt{1155/2}/16\pi^2$\\ 
$\rule[-2mm]{0cm}{5mm}$
$1$&$1$&$
(0,0,0,-\frac{1}{3},1,0)\,
            3\sqrt{77}/16\pi^2$\\ 
\hline
$\rule[-0mm]{0cm}{5mm}$
$\thalf$&$\hhalf$&$
(0,0,0,v,0,0)\,
            \sqrt{15015}/16\pi^2$\\ 
$\rule[-2mm]{0cm}{5mm}$
$\thalf$&$\thalf$&$
(-\frac{7}{11},0,1,0,0,0)\,
            11\sqrt{273/5}/32\pi^2$\\ 
\hline
$\rule[-0mm]{0cm}{5mm}$
$2$&$0$&$
(v,-\frac{1}{13},0,0,0,0)\,
            39\sqrt{77}/32\pi^2$\\ 
$\rule[-0mm]{0cm}{3mm}$
$2$&$1$&$
(0,0,0,\frac{10}{143},-\frac{11}{13},1)\,
            429\sqrt{7/86}/16\pi^2$\\ 
$\rule[-2mm]{0cm}{5mm}$
$2$&$2$&$
(0,0,0,-\frac{6}{43}+u,-\frac{22}{43},
            \frac{26}{43})\,3\sqrt{6149/2}/16\pi^2$\\ 
\hline
$\rule[-0mm]{0cm}{5mm}$
$\fhalf$&$\hhalf$&$
(0,0,0,-\frac{v}{3},v,0)\,
            3\sqrt{51051/2}/16\pi^2$\\ 
$\rule[-0mm]{0cm}{3mm}$
$\fhalf$&$\thalf$&$
(-\frac{11}{195},v,\frac{1}{15},
               0,0,0)\,39\sqrt{1785}/64\pi^2$\\ 
$\rule[-2mm]{0cm}{5mm}$
$\fhalf$&$\fhalf$&$
(-\frac{1}{4}+u,-\frac{13v}{44},
               \frac{5}{44},0,0,0)\,33\sqrt{221/7}/16\pi^2$\\ 
\hline
$\rule[-0mm]{0cm}{5mm}$
$3$&$0$&$
(-\frac{10v}{17},\frac{1}{51},v,0,
               0,0)\,51\sqrt{4389/5}/32\pi^2$\\ 
$\rule[-0mm]{0cm}{3mm}$
$3$&$1$&$
(-\frac{6v}{13},-\frac{12}{65}+u,
               \frac{38v}{65},0,0,0)\,39\sqrt{17765/7}/64\pi^2$\\ 
$\rule[-0mm]{0cm}{3mm}$
$3$&$2$&$
(0,0,0,\frac{4}{663}-\frac{u}{17}+
               v^2,0,0)\,663\sqrt{209/7}/32\pi^2$\\ 
$\rule[-2mm]{0cm}{5mm}$
$3$&$3$&$
(0,0,0,\frac{2}{51}-\frac{3u}{17},
               -\frac{6}{17}+u,\frac{4}{17})\,51\sqrt{2717/7}/32\pi^2$\\ 
\hline
$\rule[-0mm]{0cm}{5mm}$
$\shalf$&$\hhalf$&$
(0,0,0,\frac{28v}{323},
               -\frac{15v}{19},v)\,969\sqrt{231/10}/32\pi^2$\\ 
$\rule[-0mm]{0cm}{3mm}$
$\shalf$&$\thalf$&$
(0,0,0,-\frac{12v}{65}+uv,
              -\frac{6v}{13},\frac{38v}{65})\,39\sqrt{53295/2}/32\pi^2$\\ 
$\rule[-0mm]{0cm}{3mm}$
$\shalf$&$\fhalf$&$
(\frac{10}{969}-\frac{u}{19}+
               v^2,-\frac{2v}{19},-\frac{2}{323},0,0,0)\,969\sqrt{429/14}/32\pi^2
               $\\ 
$\rule[-2mm]{0cm}{5mm}$
$\shalf$&$\shalf$&$
(\frac{44}{399}-\frac{93u}{133}+
               \frac{2v^2}{7},\frac{2v}{7},-\frac{20}{133}+u,0,0,0)\,
               19\sqrt{51051/2}/64\pi^2$\\
\hline
\end{tabular}
\end{center}
\caption{
Orthonormal spherical harmonics for $L<4$
from Ref.~\ci{vanBaal:2001ac}}
\lab{tab:Yn}
\end{table}

Construction of the spherical harmonics
\begin{equation}
\cY^{(\ell,m)}=\langle u,v|{\ell,m}\rangle
=(\cY^{(\ell,m)}_1,\cY^{(\ell,m)}_2,\ldots,\cY^{(\ell,m)}_6) \,,
\end{equation}
is shown in Ref.~\ci{vanBaal:2001ac}. 
They satisfy the eigenequation
\begin{equation}
\hat{H}_{\Delta }
\cY^{(\ell,m)}
=L(2L+7)
\cY^{(\ell,m)} \,,
\lab{eq:HY}
\end{equation}
where the {\it angular momentum} 
$L=\hhalf(\ell-3)$ is a half-integer number
and its degeneration is described for {\it even} $2 L$ 
by 
\begin{equation}
m=0,1,2,\ldots,L\,,
\lab{eq:mint}
\end{equation} 
and for {\it odd} $2 L$ by
\begin{equation}
m=\frac1{2},\frac3{2},\ldots,L\,.
\lab{eq:mint}
\end{equation} 
A first few spherical harmonics are presented in Table {\ref{tab:Yn}}.
As we can see
the spherical harmonics are six-component
vectors of polynomials in $u$ and $v$.

The radial part of the Schr\"odinger equation 
for the harmonic oscillator gives the eigenequation
\begin{equation}
\left[
-\hhalf\r^{-8}\partial_\r\r^{8}\partial_\r
+\hhalf \frac{(\ell+4)(\ell-3)}{\r^{2}}
+\hhalf \r^{2}
\right]
\phi_n^{\ell}(\r)
=\tilde E_n^{\ell}\phi_n^{\ell}(\r)\,.
\lab{eq:Hrphi}
\end{equation}
The solution to Eq. (\ref{eq:Hrphi}) has a form
\begin{equation}
\phi_n^\ell(\r)= \sqrt{2 n!}\, 
\frac{ \e^{-\r^2/2}\, \r^{\ell-3}\,
L_n^{\ell+\frac12}(\r^2) }{\sqrt{\Gamma(n+\ell+\frac32})}\,,
\lab{eq:psir}
\end{equation}
where $L_n^{\ell}(x)$ 
are Laguerre polynomials, while quantum numbers
\begin{equation}
n=0,1,2,\ldots,\infty,
\lab{eq:pp}
\end{equation}
and
\begin{equation}
\ell =2 L+3=3,4,5,\ldots, \infty\,,
\lab{eq:ll}
\end{equation}
enumerate different radial solutions.
The oscillator energy defined by (\ref{eq:Hrphi}) is a function of 
the above quantum numbers:
\begin{equation}
\tilde E_n^{\ell}=\ell+2 n +\thalf.
\end{equation}
Since the number of degrees of freedom in the system is nine
a number of bosonic quanta for a given basis vector (\ref{eq:Psih2})
reads
\begin{equation}
n_B=\tilde E_n^{\ell}-\frac9{2}=\ell+2 n - 3\,.
\end{equation}
The above formula will be used to 
relate the quantum numbers $\ell$ and $n$
to the cut-off $B\ge n_B$.

Let us consider the 
integrals over $(\r,u,v)$ 
which appear in the scalar products of the wavefunctions, 
(\ref{eq:PsiPsi}) and (\ref{eq:PsiHPsi}).
Due to the simplicity of $(\r,u,v)$ variables and the polynomial form of
the eigenstates 
$\cY_{n}$ to (\ref{eq:HY}), one can perform the integrals 
of the spherical harmonics over $u$ and $v$:
\begin{equation}
X_{i,j}\equiv\int_{\r=1}d^9\c~u^iv^{2j}\,,
\lab{eq:Xint}
\end{equation}
by making use of the recurrence relation
\begin{equation}
X_{i,j}=\frac{4i(1+i+4j)X_{i-1,j}+12i(i-1)X_{i-2,j+1}
+2j(2j-1)X_{i+1,j-1}}{
(4i+6j)(4i+6j+7)}\,,
\lab{eq:Xij}
\end{equation}
where $X_{0,0}=32\pi^4/105$.
Since the radial functions (\ref{eq:psir}) 
are orthonormal
using
\begin{equation}
\int d\r~\r^8\phi_n^\ell(\r)^*\phi_{n'}^\ell(\r)=\delta_{nn'}\,,
\lab{eq:rint}
\end{equation}
one can also easily perform the integrals over $\r$.

\subsection{Hamiltonian matrix and the eigenfunctions}

In the last section we have shown the way to construct
the Hamiltonian matrix of the SYMQM model for $D=2$ and the $SU(2)$ group.
Now, we are ready to construct the matrix elements and the 
eigenfunctions of the Hamiltonian in the cut Fock space
(\ref{eq:Psih2}).
Thus, applying (\ref{eq:PsiHPsi}) and 
(\ref{eq:Xint})-(\ref{eq:rint}) we calculate
matrix elements of (\ref{eq:Hmn}) as
\begin{equation}
H^{(n',\ell',m'),(n,l,m)}=\langle n',\ell',m'|H| n,\ell,m \rangle\,,
\lab{eq:Hpnpn}
\end{equation}
where $| n,\ell,m \rangle \equiv | \Psi^{(n,\ell,m)}(\r,u,v) \rangle$.
Next, we solve the eigenequation 
\begin{equation}
\sum_{(n,\ell,m)} H^{(n',\ell',m'),(n,\ell,m)} v_k^{(n,\ell,m)}
=E_k v_k^{(n',\ell',m)}\,,
\lab{eq:Heigen}
\end{equation}
where the eigenvalues $E_k$ describe the spectrum of the Hamiltonian 
(\ref{eq:HBF})\footnote{To perform the calculation the van Baal's program \ci{vanBaal:2001ac} was rewritten
from Mathematica code to C++. This 
speeded up the program and made possible computation for higher cut-offs.}.
Finally, 
we find
the eigenfunctions of (\ref{eq:HBF}) 
in the basis (\ref{eq:Psih2})
as
\begin{multline}
|\Phi_k(\r,u,v)\rangle
=\sum_{(n,\ell,m)} v_k^{(n,\ell,m)} |\Psi^{(n,\ell,m)}(\r,u,v)\rangle=\\
=\sum_{(n,\ell,m)} v_k^{(n,\ell,m)}
\sum_{m=1}^6
\cY_{\mu}^{(\ell,m)}(u,v) \phi_n^{\ell}(\r) |e_\mu \rangle \,,
\lab{eq:Phik}
\end{multline}
where $\cY_{\mu}^{(\ell,m)}(u,v)$ 
are spherical harmonics satisfying (\ref{eq:HY})
while $\phi_n^{\ell}(\r)$ are defined  by (\ref{eq:psir}).
The non-orthogonal vectors $|e_\mu \rangle$ contain
the fermionic variables and their scalar product is defined
by (\ref{eq:Nnm}).
The above formula is the main expression which we use to 
compute the wave-functions of the model.

\section{Results}
\subsection{Eigenenergies, coexistence and $B-$dependence}

Solving the eigenequation (\ref{eq:Heigen})
for different cut-off $B$ 
we obtain the energy spectrum of the system with $j=0$ and $n_F=2$
as a function of $B$. 
Using Alpha DEC, PC computers and optimized C++ code
we were able to reach the cut-off values $B=60$. 
The spectrum is shown for $B \le 60$
in Fig.~\ref{fig:Eb}.
\begin{figure}[ht!]
\centerline{
\begin{picture}(130,80)
\put(2,70){$E$}
\put(2,0){\epsfysize7.6cm \epsfbox{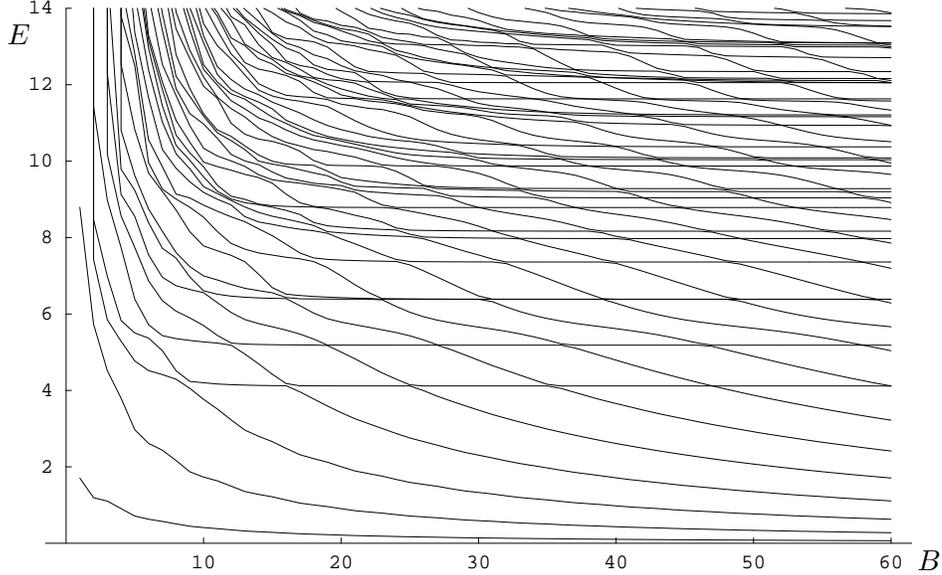}}
\put(123,0){$B$}
\end{picture}
}
\caption{The energy spectrum as a function of
cut-off $B\ge n_B$}
\lab{fig:Eb}
\end{figure}
Looking at this Figure one can  notice
two kinds of behaviour.
The levels from the first group are the ones which are
rapidly, possibly exponentially, convergent to
finite energy values
and they become constant for large $B$.
These values represent the full non-perturbative
eigenspectrum of the ``un-cut'' system.
The other curves fall down slowly as 
\begin{equation}
E_k(B) \sim \frac{1}{B}\,.
\lab{eq:E1B}
\end{equation}

These two kinds of curves do 
not cross but one can see, especially for lower $B$,
that the levels repel each other exchanging their own $B$-behaviour. 
For higher $B$ this repulsion
is more subtle.

There is a conjecture
that the asymptotics of the wave function 
at large distances, \ie large $\r$, determine the convergence of our 
calculations being performed with the 
enlarging number of allowed quanta $B$ \ci{Wosiek:2002nm}.
Thus, localized states, whose wave-functions do not go
deep into the valleys, converge faster with the cut-off.
These states are related to the discrete spectrum of the 
Hamiltonian (\ref{eq:HBF}).
They are supersymmetric partners of the
the localized states from $n_F=1,3$ sectors from supersymmetric multiplets
\ci{Wosiek:2002nm,Campostrini:2004bs}.
To obtain spectrum of these states
the limit
\begin{equation}
E=\lim_{B \to \infty} E_{B,k} \Big|_{k=\mbox{const}}\,,
\lab{eq:dislim}
\end{equation}
is performed
where index $k$ enumerates consecutive energy curves
and changes only at the anti-crossing points.

On the other hand,
the non-localized states penetrate the valleys
with increasing the cut-off.
They have the power-like behaviour of the energy level.
The energy curves which fall slowly to the zero energy with
increasing cut-off never converge.
They form the continuous spectrum at $B=\infty$.
It was in Ref.~\ci{Trzetrzelewski:2003sz} 
shown that non-trivial and correct continuum limit
for these states is given by
\begin{equation}
E(P)=\lim_{B \to \infty} E_{B,k(B,P)} \Big|_{P=\mbox{const}}
\quad
\mbox{with}
\quad
k(B,P)=\frac{P}{\pi}\sqrt{2 B}\,
\lab{eq:Binf}
\end{equation}
where $P$ is the continuous momentum, $B$ is the cut-off
and $k$ enumerates only energy levels from the continuous spectrum. 
The continuous spectrum consists of
all positive energy values. 
Localized and non-localized states coexist
as a consequence of the supersymmetric interactions with flat valleys.

It turns out that 
the vacuum state belongs to the sector with $j=0$
and $n_F=2$ and is formed
by the continuous spectrum
\ci{vanBaal:2001ac,Wosiek:2002nm}. 
In order to obtain this state
one has to take the states from the continuous
spectrum, perform the continuous limit (\ref{eq:Binf})
and at the end go with $E \to 0$.
Since the SUSY vacuum is constructed from the states
with continuous energy spectrum,
it is non-normalizable
\footnote{The vacuum state is non-normalizable
means its {\it normalization} to the Dirac delta function.} 
\ci{Wosiek:2002nm}.

\begin{figure}[ht!]
\centerline{
\begin{picture}(120,80)
\put(0,65){$E$}
\put(2,0){\epsfysize7.5cm \epsfbox{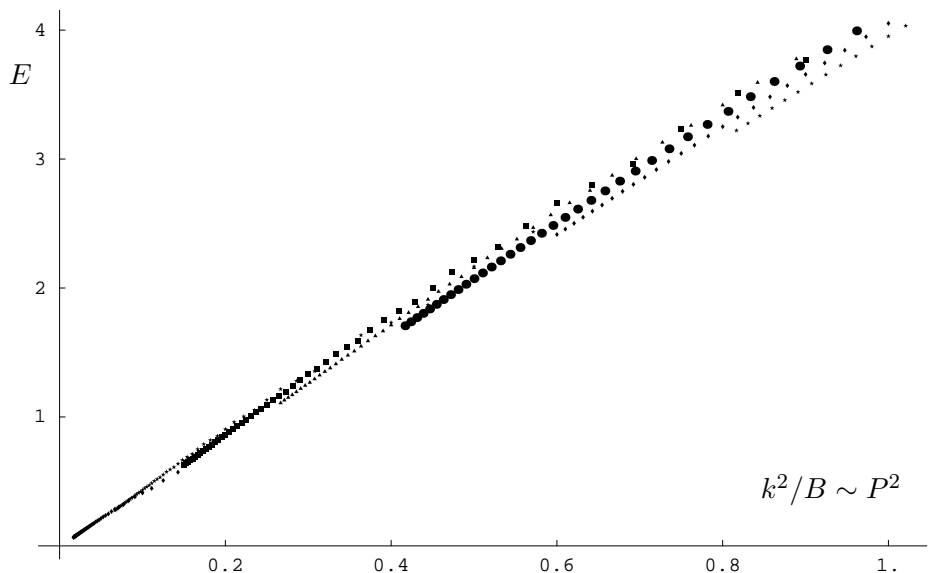}}
\put(100,10){$k^2/B \sim P^2$}
\end{picture}
}
\caption[Dispersion relation: $E\sim P^2$.]
{Dispersion relation, $E\sim P^2$, for a few first
energy curves from the continuous spectrum. Here, 
$k$ - enumerates consecutive energy curves for the continuous spectrum and
according to \ci{Trzetrzelewski:2003sz} $k$
is related to momentum  $P$ as $k/\sqrt{B} \sim P$. }
\lab{fig:Ereldis}
\end{figure}

\subsection{Scaling and dispersion relation}

It was shown in Ref.~\ci{Trzetrzelewski:2003sz}
that for the free Hamiltonian 
the momentum of its eigenstate $P$ is related to 
a discrete index of consecutive continuous-energy levels $k$
by 
\begin{equation}
P \sim \frac{k}{\sqrt{B}}\,,
\lab{eq:rd}
\end{equation} 
where $B$ is a number of bosons.
It turns out that similar dispersion relation holds
for our system. 
This is seen in Fig.~\ref{fig:Ereldis}
where $E(P^2)$ is plotted.
In this Figure the energies from a distinct energy curves 
of continuous spectrum, enumerated by $k$, 
are denoted by different symbols.
In order to evaluate $P$ Eq.~(\ref{eq:rd}) was used. 
Indeed, the points in the plot form a curve $E(P) \sim P^2$
with a  very good approximation. 
Thus, one can
suppose that these states form the continuous spectrum at $B=\infty$.
This again agree with the continuous nature of scattering states.
Therefore, 
we confirm that 
the states which correspond to the energy levels with power-like behaviour 
describe asymptotically free particles.
One can imagine that these particles
propagate in flat valleys of 
the potential (\ref{eq:Vx}). 

The above scaling law (\ref{eq:Binf})
is required to recover the infinite Hilbert space 
limit (\ref{eq:Binf}).
On the other hand it is interesting because its universality.
It is valid for models with only continuous spectrum as well 
as the mixed one, 
like the one considered in this work
where the Hamiltonians are less trivial
and the localized and non-localized states coexist at the same 
energy.

\subsection{Localized versus non-localized eigenfunctions}

Applying (\ref{eq:Phik}) one can also calculate the eigenstates of 
(\ref{eq:Hmn}). 
A simple way to analyse the bosonic probability density
is to use
\begin{equation}
|\Phi_k(x_1,x_2,x_3)|^2 \equiv
\langle
\Phi_k(\r,u,v)
| \Phi_k(\r,u,v)
\rangle_F\;,
\lab{eq:dens}
\end{equation}
where we apply (\ref{eq:Nnm}) while the integration in this product 
is only performed over the fermionic degrees of freedom.

\begin{figure}[ht!]
\centerline{
\begin{picture}(120,75)
\put(115,0){$x$}
\put(-5,70){$|\Phi_{s=1}^{(L)}(x)|^2$}
\put(90,48){\fbox{$s=1$}}
\put(0,0){\epsfysize 8cm \epsfbox{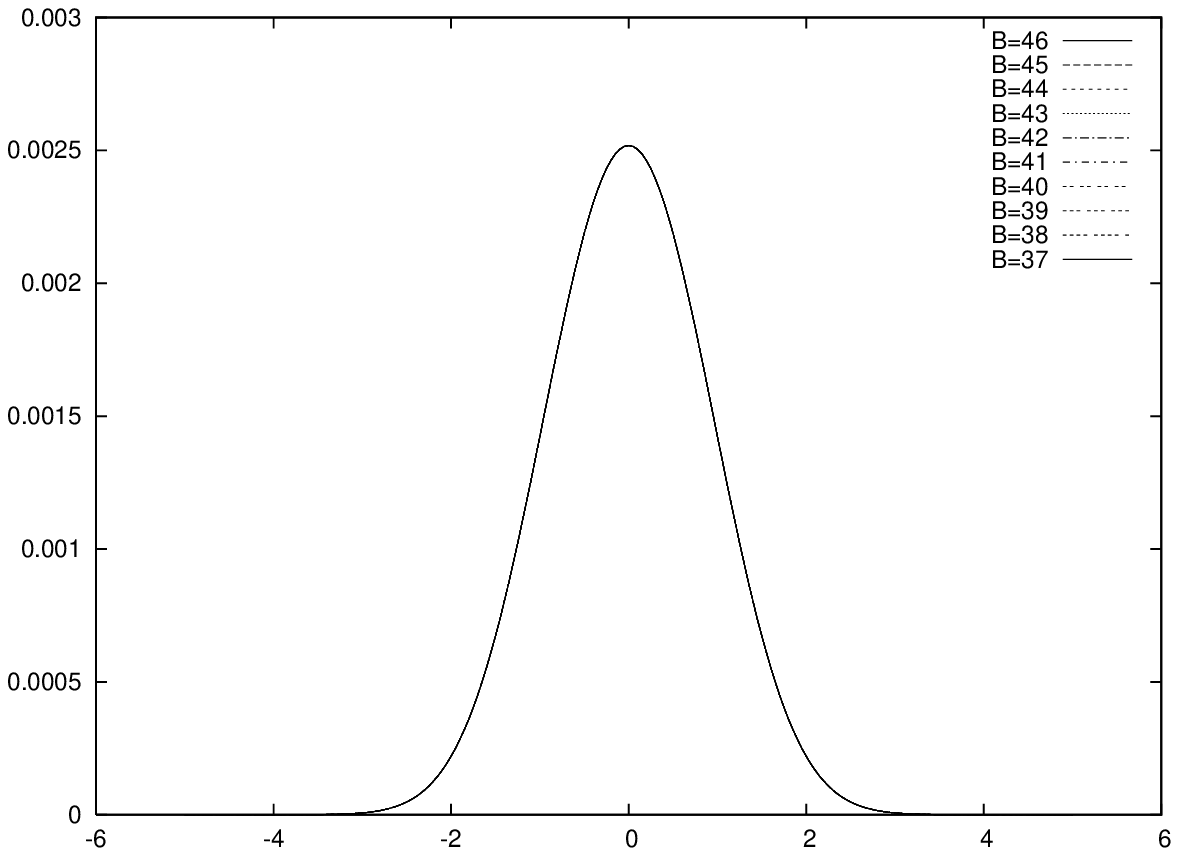}}
\end{picture}}
\caption{The first bound eigenstate $s=1$ for $B=37,\ldots,46$:\newline
$|\Phi_{s=1}^{(L)}(x,x_2=0,x_3=0)|^2$}
\lab{fig:bs1}
\end{figure}

\begin{figure}[ht!]
\vspace{0.5cm}
\centerline{
\begin{picture}(120,75)
\put(115,0){$x$}
\put(-5,70){$|\Phi_{s=2}^{(L)}(x)|^2$}
\put(90,48){\fbox{$s=2$}}
\put(0,0){\epsfysize 8cm \epsfbox{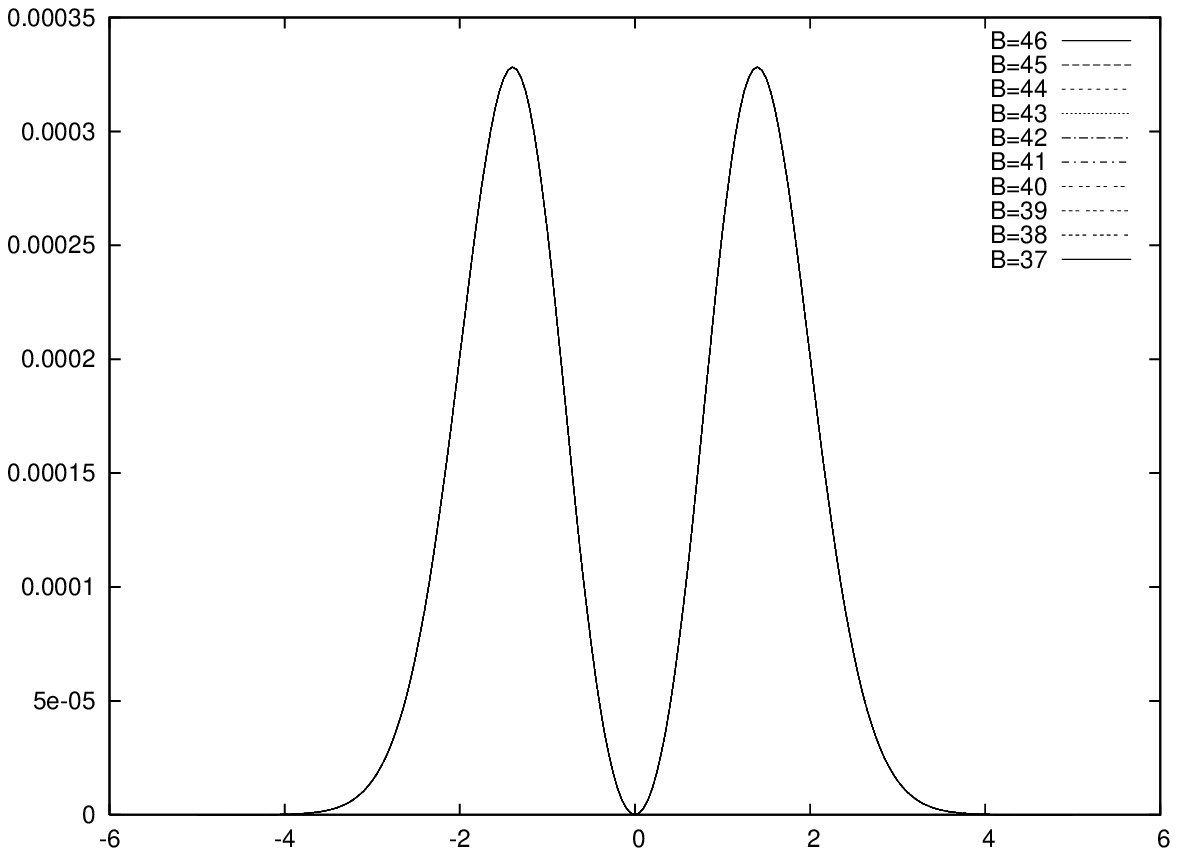}}
\end{picture}
}
\caption
{This is Fig.~\ref{fig:bs1} but the bound eigenstate with 
$s=2$ for $B=37,\ldots,46$:\newline
$|\Phi_{s=2}^{(L)}(x,x_2=0,x_3=0)|^2$
}
\lab{fig:bs2}
\end{figure}
\begin{figure}[ht!]
\centerline{
\begin{picture}(120,75)
\put(115,0){$x$}
\put(-7,70){$|\Phi_{s=3}^{(L)}(x)|^2$}
\put(90,48){\fbox{$s=3$}}
\put(0,0){\epsfysize 8cm \epsfbox{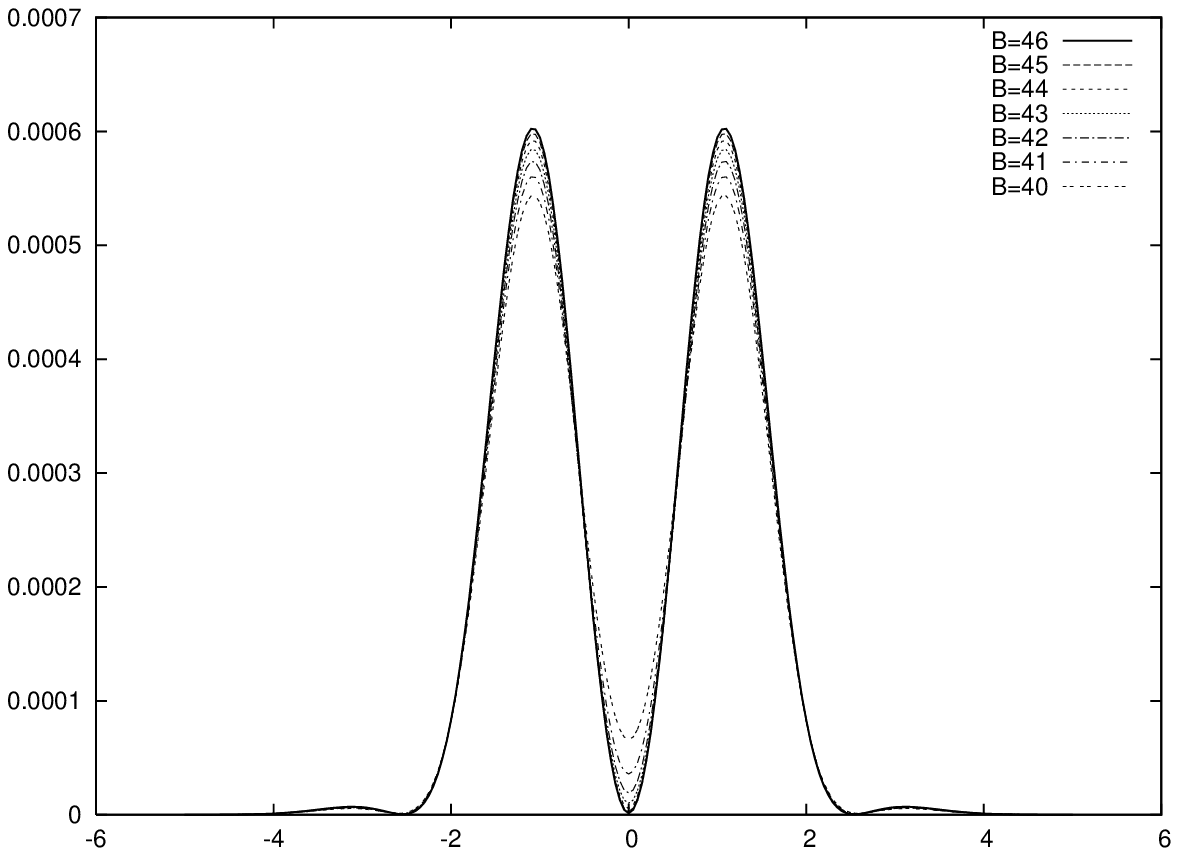}}
\end{picture}
}
\caption
{Same for the third bound eigenstate $s=3$ for $B=40,\ldots,46$:\newline
$|\Phi_{s=3}^{(L)}(x,x_2=0,x_3=0)|^2$}
\lab{fig:bs3}
\end{figure}
\begin{figure}[ht!]
\centerline{
\begin{picture}(120,75)
\put(115,0){$x$}
\put(-9,72){$|\Phi_{s=4}^{(L)}(x)|^2$}
\put(90,48){\fbox{$s=4$}}
\put(0,0){\epsfysize 8cm \epsfbox{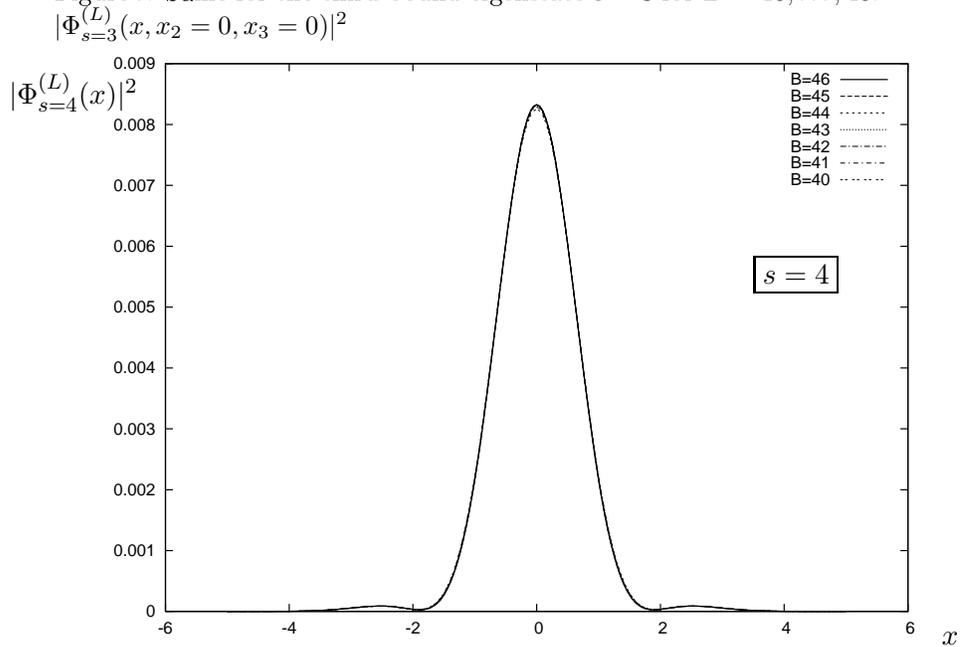}}
\end{picture}
}
\caption
{And for the fourth bound eigenstate $s=4$ for $B=40,\ldots,46$:\newline
$|\Phi_{s=4}^{(L)}(x,x_2=0,x_3=0)|^2$
}
\lab{fig:bs4}
\end{figure}
\begin{figure}[ht!]
\centerline{
\begin{picture}(120,70)
\put(0,0){\epsfysize6.5cm \epsfbox{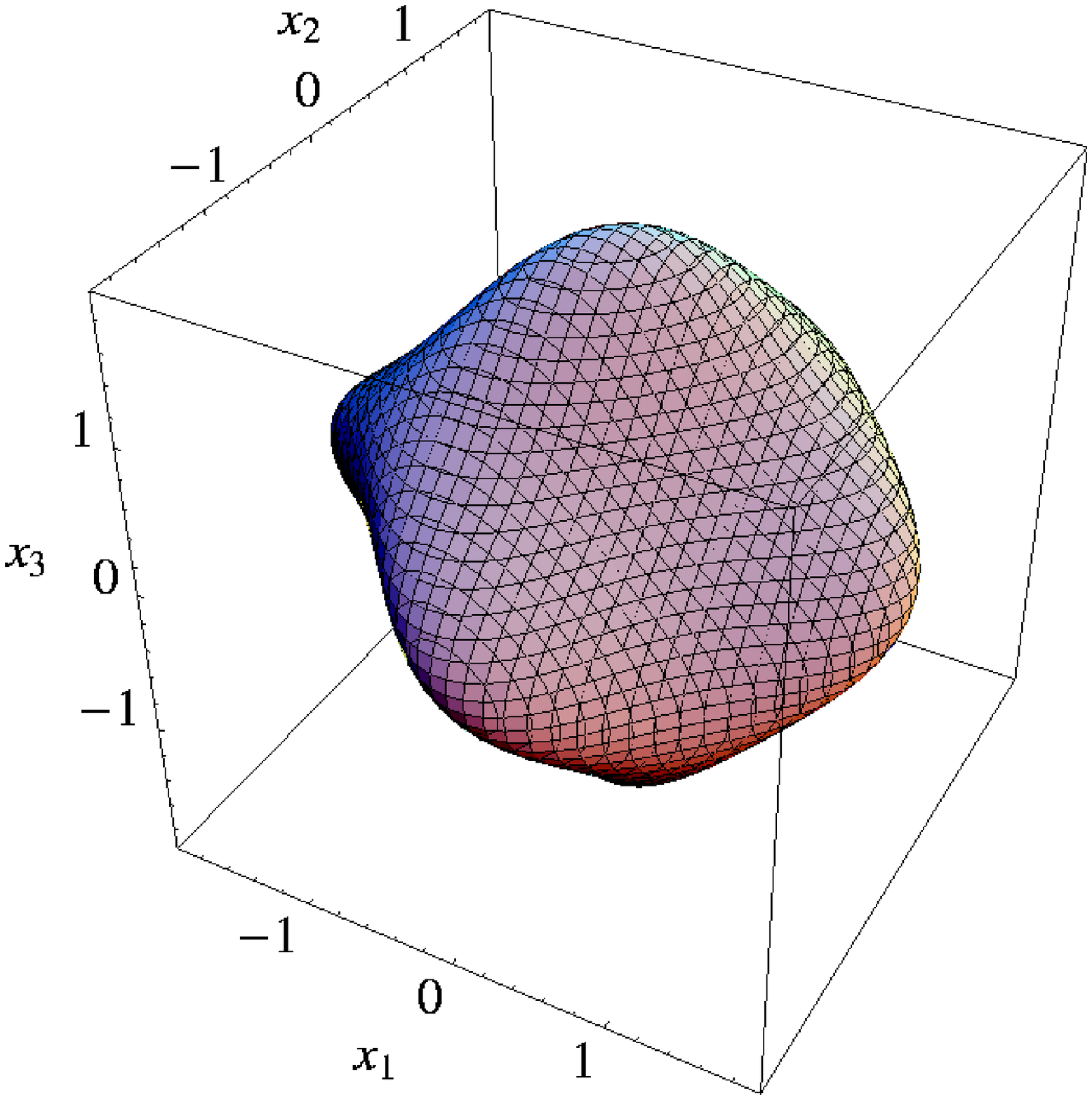}}
\put(60,0){\epsfysize6.5cm \epsfbox{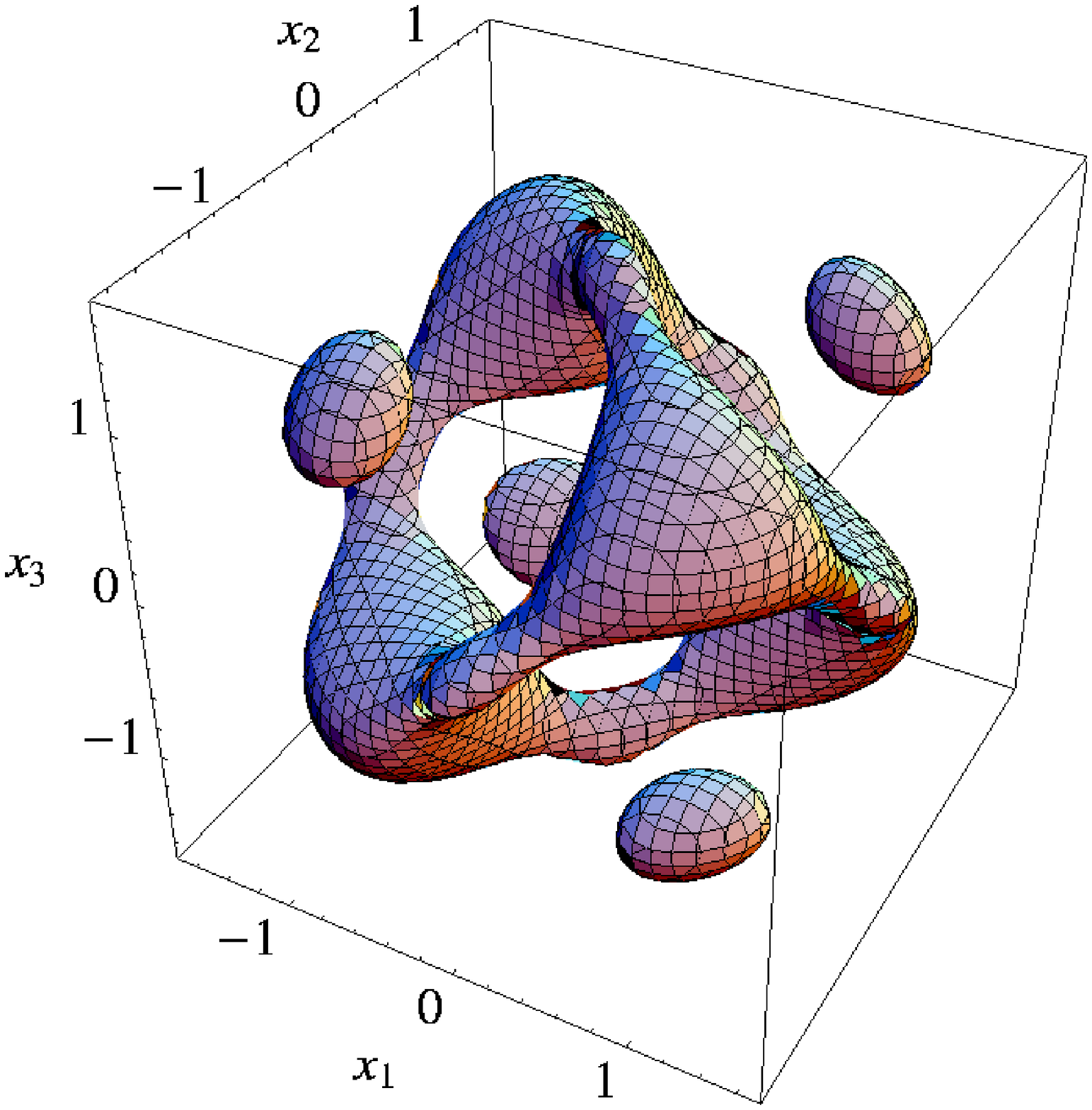}}
\end{picture}
}
\caption{Contour plots for the bound eigenstates with $s=1,2$ for
$|\Phi_{s}^{(L)}(x_1,x_2,x_3)|^2=0.0003$
}
\lab{fig:s3d12}
\end{figure}

The energy levels of localized states
converge rapidly with $B$.
Similarly, their eigenstates become practically $B$ independent.
As one can see 
due to anti-crossing energy levels
(see Fig.~\ref{fig:Eb}) the localized states with a specified
energy (but considered for different cut-off $B$) corresponds
to distinct values of $k$.
For our convenience
we denote the localized states 
by index $(L)$, \ie $\Phi_s^{(L)}(\r,u,v)$,
where contrary to
$\Phi_k(\r,u,v)$ case, 
index $s$ enumerates only states of the discrete spectrum.
A few first bound states are shown in Figs.~\ref{fig:bs1}-\ref{fig:bs4}.
There are mainly localized around the origin where potential has wide
minimum -- the stadium.
These Figures show plots with various cut-off's $B$.
The convergence of the first two states is so good that it is not possible to 
distinguish the wave-functions with different cut-off's.
For higher states with $s=3$, and  $s=4$
this convergence is worse\footnote{Because of the simpler shape,
the wave-function with $s=4$ converges more rapidly than one with $s=3$.
However, this convergence is much slower than one for the other
wave-functions, \ie with $s=1, 2, 5, \ldots\,$.}. 
This is caused by interference of the states 
which have 
nearly the same  energy (see Fig.~\ref{fig:Eb}).
The next state with $s=5$ is not degenerate
and its wave-function also rapidly converges.

In order to observe the symmetry of the localized states we present also
contour plot for 
$|\Phi_{s}^{(L)}(x_1,x_2,x_3)|^2=0.003$ in Figs.~\ref{fig:s3d12}-\ref{fig:s3d34}.
One can see that only the first
bound state possesses the symmetry of the 
bosonic potential (\ref{eq:Vx}):
\begin{equation}
x_i \leftrightarrow \pm x_j \quad \mbox{for }  i,j=1,2,3.
\lab{eq:symxixj}
\end{equation}
For the remaining states one part of the symmetry (\ref{eq:symxixj}),
\ie (\ref{eq:xtmx}),
can be broken by the fermionic term in the Hamiltonian (\ref{eq:HFmat})
\footnote{$H_F$ contains $v$ variables without square powers.
This breaks the symmetry (\ref{eq:xtmx}) of the bosonic potential.}.
Thus, in the fermionic case the symmetry (\ref{eq:xtmx})
is not valid.
Although in the considered case $j=0$
the eigenstates are not spherically symmetric
and they have complicated shapes of contour plots 
with a genus sometimes greater than zero.
However, one has to remember
that the angular momentum $j=0$ applies for
the $9-$dimensional space of ${\c^a_i}$.
Since the transformation (\ref{eq:cdiag})
of the $\c-$coordinate space to the 
$3-$dimensional $x_i$-space
is non-linear
the three-dimensional spherical symmetry in the $x_i$-space
is not obvious.

\begin{figure}[ht!]
\centerline{
\begin{picture}(120,70)
\put(0,0){\epsfysize6.5cm \epsfbox{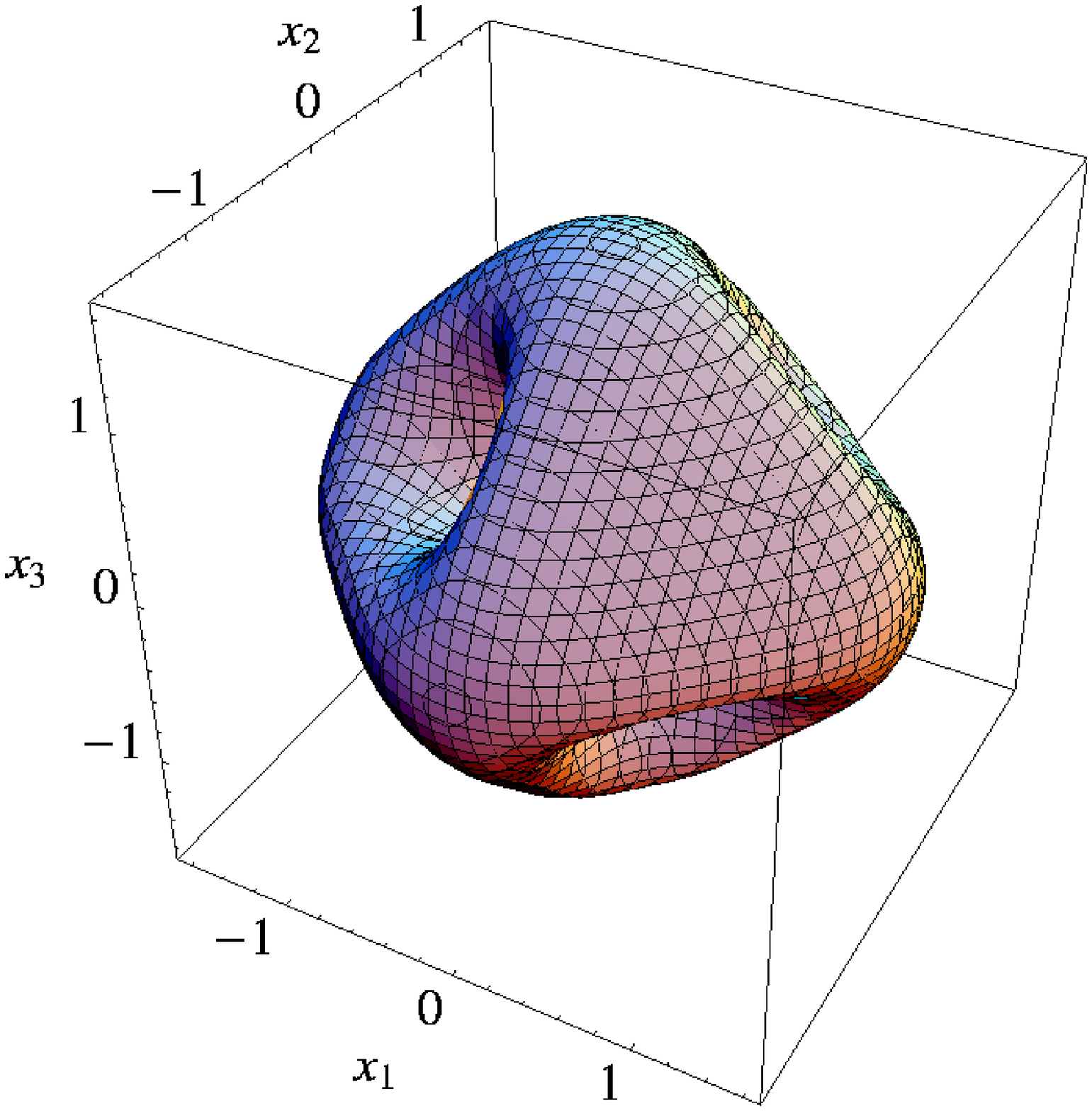}}
\put(60,0){\epsfysize6.5cm \epsfbox{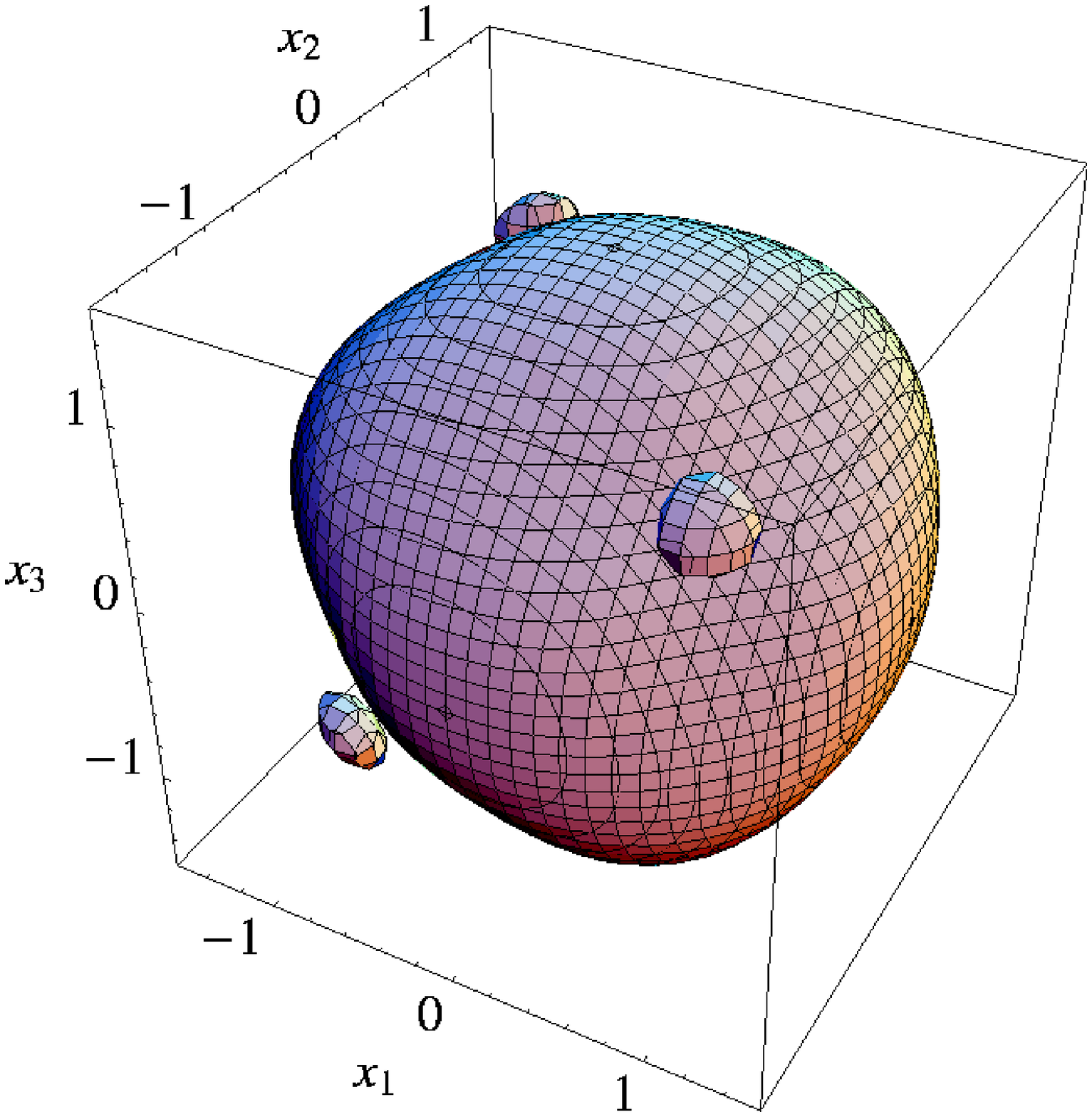}}
\end{picture}
}
\caption{The same contour plots but for the bound eigenstates 
with $s=3,4$ for
$|\Phi_{s}^{(L)}(x_1,x_2,x_3)|^2=0.0003$
}
\lab{fig:s3d34}
\end{figure}

\begin{figure}[ht!]
\centerline{
\begin{picture}(120,75)
\put(115,0){$x$}
\put(0,72){$x^2 |\Phi_k^{(NL)}(x)|^2$}
\put(100,53){\fbox{$E=1.7$}}
\put(10,0){\epsfysize8cm \epsfbox{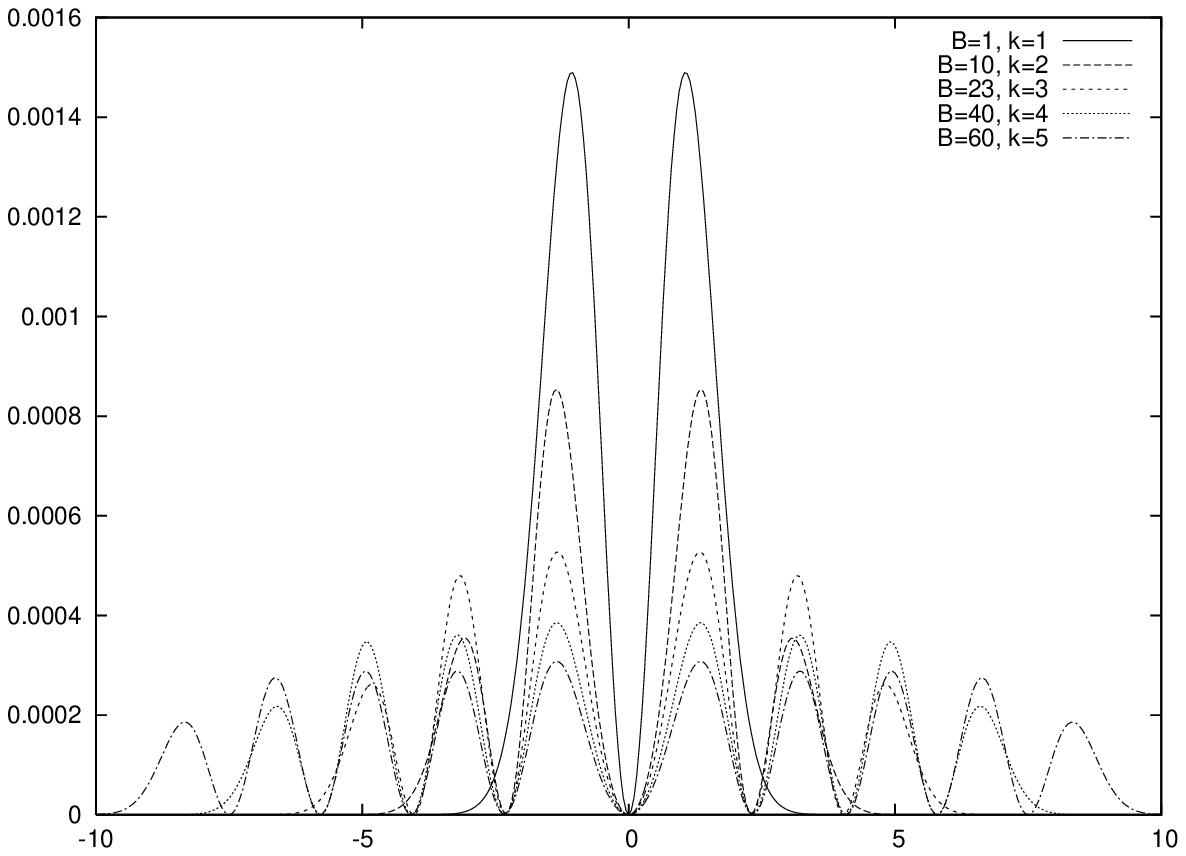}}
\end{picture}
}
\caption
{The functions $x^2 |\Phi_k^{(NL)}(x)|^2$ for eigenstates with energy levels 
$E_k(B)$ from the continuous spectrum.
Here $E= 1.7$ and $k=1,2,3,4,5$.}
\lab{fig:E1_7}
\end{figure}
\begin{figure}[ht!]
\vspace{0.5cm}
\centerline{
\begin{picture}(120,75)
\put(115,0){$x$}
\put(0,73){$x^2 |\Phi_k^{(NL)}(x)|^2$}
\put(100,53){\fbox{$E= 2.6$}}
\put(10,0){\epsfysize8cm \epsfbox{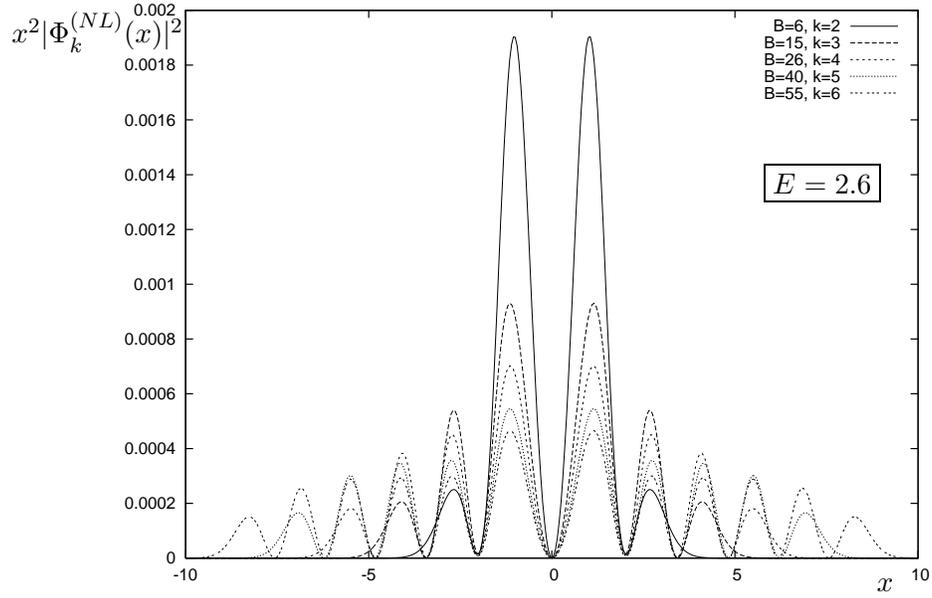}}
\end{picture}
}
\caption
{The plot similar to ones from Fig.~\ref{fig:E1_7} 
but for the eigenstates with $E=2.6$ 
and $k=1,2,3,4,5$.}
\lab{fig:E2_6}
\end{figure}
\begin{figure}[ht!]
\vspace{0.5cm}
\centerline{
\begin{picture}(120,75)
\put(115,0){$x$}
\put(0,71){$x^2 |\Phi_k^{(NL)}(x)|^2$}
\put(100,53){\fbox{$E= 3.7$}}
\put(10,0){\epsfysize8cm \epsfbox{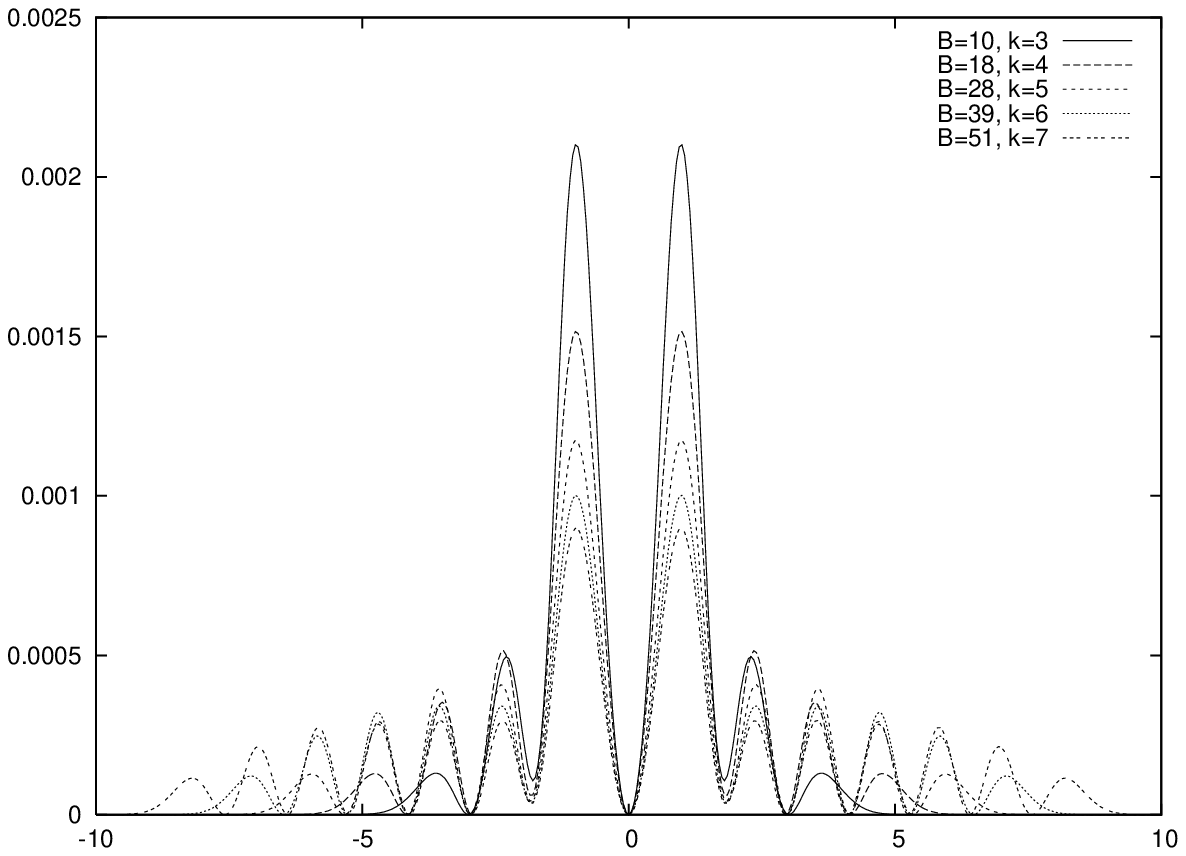}}
\end{picture}
}
\caption
{The same function as in Fig.~\ref{fig:E1_7}  but for $E=3.7$ 
and $k=1,2,3,4,5$.}
\lab{fig:E3_7}
\end{figure}

Let us consider the probability density of 
states with continuous spectrum along 
the potential valley:
\begin{equation}
x_1=x, \quad x_2=x_3=0.
\lab{eq:val}
\end{equation}
It turns out that going away from $x_1=x_2=x_3=0$
the amplitude of the eigenstates (\ref{eq:dens})
is strongly suppressed. Thus,
to see its dependence along the valley
a quantity 
\begin{equation}
x^2 |\Phi_k^{(NL)}(x)|^2 \equiv
x^2 |\Phi_k^{(NL)}(x_1=x,x_2=0,x_3=0)|^2
\lab{eq:xPhi}
\end{equation}
is studied.
Similarly to the localized states
the non-localized ones are denoted by
superscript $(NL)$, \ie
$\Phi_k^{(NL)}(x_1,x_2,x_3)$, and
their index $k$ enumerates only the energy levels
from the continuous spectrum.  
The scaling limit implied in (\ref{eq:Binf})
is performed as follows.
Take an arbitrary energy, i.e $E=1.7$,
and for a given energy curve 
from the continuous spectrum, \ie  $E^{(NL)}_k(B)$
enumerated by index $k$,
find the cut-off  $B$
which best satisfies the following condition
\begin{equation}
E^{(NL)}_k(B)=E\,,
\lab{eq:Eb=E}
\end{equation}
where $E$ is the arbitrary energy.
The probability density
of the Hamiltonian eigenstates
multiplied by $x^2$ (\ref{eq:xPhi})
is shown for $E=1.7$ and $k=1,2,3,4,5$
in Fig.~\ref{fig:E1_7}.
As one can see for $k=1$ 
the function $x^2 |\Phi_k^{(NL)}(x)|^2$
has two peaks. For the next energy curve,
\ie $k=2$, the plot has four peaks with lower amplitudes.
Thus for the $k-$th curve one can see $2 k$ peaks. 
In the limit (\ref{eq:Binf}) the amplitudes of the peaks for a given energy
are nearly the same.
This does not occur for the peaks in the centre of the potential
and the peaks deeper sunk in the valley.
For all $k$ the oscillations of frequencies
are nearly the same and correspond to the energy $E=1.7$.
For higher energies, \ie $E=2.6$ and $E=3.7$ presented
in Fig.~\ref{fig:E2_6} and Fig.~\ref{fig:E3_7} respectively,
frequencies are lower but
for the fixed energy value they are approximately independent of $k$.
Taking states for higher energy curves
one can notice that 
their wave-functions  enter
deeper into the valleys (\ref{eq:Vx}).
This fact confirms that the 
considered states
form the continuous spectrum for $B\to\infty$.

In the purely bosonic sector, \ie where $n_F=0$, $6$,
as well as in the sectors with $n_F=1,5$, where bosonic modes are dominating,
despite the flatness of the potential valleys
the flat directions are blocked by the energies of the transverse quantum 
fluctuations.
This makes the spectrum in zero-fermion and one-fermion sectors discrete.
However, in the supersymmetric system
the transverse fluctuations cancel among bosonic and fermionic states.
Thus, valleys are not blocked and in the model 
with supersymmetric fermions the continuous spectrum
appears. This occurs in sectors for $n_F=2,3,4$.

On the other hand, we have sectors with only discrete spectra, \ie 
$n_F=0,1,5,6$.
The supersymmetry creates super-multiplets and
demands the existence of similar states in others sectors.
Therefore in the central sectors, \ie $n_F=2,3,4$, there is 
coexistence of discrete and continuous spectra.
This is exactly what Fig.~\ref{fig:Eb} shows.

\section{Summary}
Despite the fact that 
Supersymmetric Yang-Mills Quantum Mechanics (SYMQM) models
\ci{Witten:1981nf,Claudson:1984th} 
are simpler than the original field theories,
they still pose 
difficult to solve complex problems. 
On the other hand we have the BFSS equivalence hypothesis 
\ci{Banks:1996vh,Bigatti:1997jy,Taylor:2001vb}, 
which relates SYMQM to $M-$theory.
This makes
SYMQM even more interesting.
Moreover, the SYMQM models form very good laboratory for learning
of supersymmetric theories.
Studies 
in their zero-fermion sectors,
\ie of the non-supersymmetric Yang-Mills mechanics,
give us knowledge about
0-volume glueball states, 
\ci{Luscher:1982ma,Luscher:1983gm,vanBaal:1988qm,Wosiek:2002nm}.
On the other hand
testing relations between fermionic sectors we
can observe specific properties of 
the action of the supersymmetric operator.

In this work we have considered the 
$D=4$ dimensional model with $SU(2)$ gauge groups.  This model is 
non-trivial and  possesses both localized and non-localized states. 
There were many approaches  \ci{Wosiek:2002nm,Kotanski:2002fz,
vanBaal:2001ac,Campostrini:2004bs} to understand the model and to 
find its energy spectrum. 
We have followed the method \ci{vanBaal:2001ac} proposed by van Baal which
allows to solve problem in the sector with specified not
only the number of fermionic quanta $n_F$ 
but also at the fixed momentum $j=0$.
We have chosen the sector with $n_F=2$ and $j=0$, where the 
supersymmetric vacuum should appear.

First, the energy spectrum for very high cut-offs 
$B\le 60$ has been calculated. It exhibits complex structure of the states,
 \ie coexistence of localized and non-localized states.
For a high cut-offs, \ie $B>40$, not only the energies of localized states
become $B-$independent
but also the corresponding eigenfunctions.
We have constructed the first few bound state and described their 
properties.
Our spectrum agree with the previous results \ci{Wosiek:2002nm,Kotanski:2002fz,
vanBaal:2001ac,Campostrini:2004bs}. However, 
with the higher cut-off it is much more accurate.

In the plot of the energy spectrum as a function of the cut-off $B$
in Fig.~\ref{fig:Eb}
the non-localized states form curves behaving as $1/B$. 
Taking the states from these curves for a constant energy 
we have  studied the properties of the eigenfunction from 
the continuous spectrum. At the end we have tested the
scaling of the continuous spectrum 
(\ref{eq:Binf})
and confirmed numerically 
dispersion relation of the free particles 
presented in Ref.~\ci{vanBaal:1988qm}. 
The scaling was derived 
in Ref.~\ci{Trzetrzelewski:2003sz} for models with
only non-localized states whereas here we have shown that with 
a very good approximation is also valid for 
the non-trivial interactions.

\section{Acknowledgements}

I thank Jacek Wosiek for 
suggesting the subject and fruitful discussions
and Pierre van Baal for
making the program for calculation the energy spectrum available for me. 
I also acknowledge
discussions with Maciej Trzetrzelewski.
This work was supported by the 
grant of the Polish Ministry of Science and Education
P03B 024 27 (2004-2007).

\end{document}